\documentclass[9pt,shortpaper,twoside,web]{ieeecolor}
\usepackage{generic}
\usepackage{cite}
\usepackage{amsmath,amssymb,amsfonts}

\usepackage{algorithm}
\usepackage[pdftex]{graphicx}
\usepackage{epstopdf}

\usepackage{algorithmic}
\usepackage{textcomp}
\def\BibTeX{{\rm B\kern-.05em{\sc i\kern-.025em b}\kern-.08em
    T\kern-.1667em\lower.7ex\hbox{E}\kern-.125emX}}
\begin{document}

\title{Fully-Heterogeneous Containment Control of a Network of Leader-Follower Systems}

\author{Majid Mazouchi, Farzaneh Tatari, Bahare Kiumarsi \IEEEmembership{Member, IEEE}, and Hamidreza Modares \IEEEmembership{Senior Member, IEEE}
\thanks{ M. Mazouchi and H. Modares are with the Department of Mechanical Engineering, Michigan State University, East Lansing, MI, 48824, USA (e-mails: \{Mazouchi,modaresh\}@msu.edu). }
\thanks{F. Tatari, is with the KIOS Research and Innovation Center of Excellence, Department of Electrical and Computer Engineering, University of Cyprus, Nicosia, 1678, Cyprus (e-mail: tatari.farzaneh@ucy.ac.cy).}
\thanks{B. Kiumarsi is with 
the Department of Electrical and Computer Engineering, Michigan State University, East Lansing, MI, 48824, USA (e-mail: kiumarsi@msu.edu).}}

\maketitle

\begin{abstract}
This paper develops a distributed solution to the fully-heterogeneous containment control problem (CCP), for which not only the followers' dynamics but also the leaders' dynamics are non-identical. A novel formulation of the fully-heterogeneous CCP is first presented in which each follower constructs its virtual exo-system. To build these virtual exo-systems by followers, a novel distributed algorithm is developed to calculate the so-called normalized level of influences (NLIs) of all leaders on each follower and a novel adaptive distributed observer is designed to estimate the dynamics and states of all leaders that have an influence on each follower. Then, a distributed control protocol is proposed based on the cooperative output regulation framework, utilizing this virtual exo-system.  Based on estimations of leaders' dynamics and states and NLIs of leaders on each follower, the solutions of the so-called linear regulator equations are calculated in a distributed manner, and consequently, a distributed control protocol is designed for solving the output containment problem. Finally, theoretical results are verified by performing numerical simulations.
\end{abstract}


\section{Introduction}
\label{sec:introduction}
\IEEEPARstart{D}{istributed} cooperative control of multi-agent systems (MASs) has gained significant interest from diverse communities, due to its numerous applications in variety of disciplines, such as cooperative networked mobile robots control, distributed sensor networks, satellite formation flying, cooperative vehicles formation control, cooperative surveillance, and so forth \cite{ji2008containment,ren2007information}. One major line of research on distributed cooperative control of MASs is the consensus or synchronization problem, which has been well investigated in the literature \cite{ren2007information}. The consensus problem is generally categorized into two main classes, namely, the leaderless consensus problem and the leader-following consensus problem. In both of these classes, agents must reach a common value or trajectory of interest, and this common value is dictated by the leader in the leader-following case. In many applications of MASs, however, agents are not meant to reach the same value or follow the same trajectory. For instance, in the containment control problem (CCP), which is the problem of interest in this work, there exist multiple leaders, and the main objective is to design distributed control protocols under which the followers are driven into the convex geometric space, i.e., convex hull, spanned by the leaders. This convex hull can, for example, provide a safe region for the followers to fall into. There has been a large body of work on the CCP \cite{cao2012distributed,ji2008containment,liu2012necessary,li2011distributed,li2013distributed,meng2010distributed,xiao2018containment,yoo2013distributed,yuan2021cooperative,jiang2018fully}, most of which considered the homogeneous CCP, in the sense that the followers' dynamics and the leaders' dynamics are all identical.

Semi-heterogeneous CCP, for which the dynamics of the leaders are identical but the dynamics of the followers can be non-identical, has been also considered in the literature. The cooperative output regulation framework has been widely employed \cite{chu2016distributed,xiao2018containment,jiang2018fully,haghshenas2015containment,huang2016certainty} to solve the semi-heterogeneous CCP. Agents' dynamics are assumed linear in \cite{chu2016distributed,haghshenas2015containment}. These results assume that the leaders' dynamics are all identical. However, in reality, the leaders' dynamics might not be identical, even for the same type of systems, or can even change over time, due to aging or component failure. 
Despite its importance, only a few results have investigated the fully heterogeneous containment control problem (FHCCP) with heterogeneous leaders and followers \cite{wang2018output,wang2017output}. In these approaches, however, it is assumed that the leaders exchange information with each other and can cooperatively design a so-called virtual leader to describe the trajectories of all followers, and the communication graph is assumed undirected. 
In \cite{li2017containment} and \cite{wang2014adaptive}, by using a recursive stabilization control method and based on the adaptive internal model approach, an adaptive control protocol is presented to solve the FHCCP for a class of linear MASs. These methods and are restricted to agents with minimum phase dynamics and the same relative degree. Moreover, existing methods presented in \cite{wang2018output,wang2017output,wang2014adaptive,li2017containment} are not fully distributed due to the requirement of knowing the Laplacian matrix of the communication graph in their design procedure.

To the best of our knowledge, the development of a fully distributed solution to the FHCCP has not been investigated in the literature, which motivates this study. Similar to \cite{li2017containment} and \cite{wang2014adaptive}, we first convert the output CCP into a set of decoupled reference trajectory tracking problems, in which each follower aims to track its virtual reference trajectory, generated by its virtual exo-system. To construct such virtual exo-systems for followers, it is first shown that each follower needs to know the dynamics of all of its influential leaders (i.e., the leaders that have at least one directed path of some length to it), as well as their so-called normalized levels of influences (NLIs). Then, a novel distributed algorithm is developed to calculate the NLIs of all leaders on each follower in a fully distributed fashion by introducing novel concepts of local adjacency matrix, local in-degree matrix, and local Laplacian matrix. To estimate the dynamics of the influential leaders that the agents are reachable from in a distributed manner, an adaptive distributed observer is designed. The proposed distributed observer also estimates the states of the leaders with nonzero NLIs. Using calculated NLIs and the estimated dynamics and states of the influential leaders on each follower, the followers construct their virtual exo-systems and calculates their control protocols in a completely distributed fashion.

\textbf{Notations:} Throughout the paper, ${\Re ^n}$ and ${\Re ^{n \times m}}$ represent, respectively, the $n-$dimensional real vector space, and the $n \times m$ real matrix space. $\mathbb{N}$  denotes the set of natural numbers. ${0_{m \times n}}$ and ${0_n}$ denote, respectively, the $m \times n$ matrix and the column vector with all entries equal to zero. Let ${1_n}$   be the column vector with all entries equal to one. ${I_n}$ represents the $n \times n$  identity matrix. $diag\left( {{d_1},...,{d_n}} \right)$ represents a block-diagonal matrix with matrices ${d_1},...,{d_n}$ on its diagonal. ${\left\| . \right\|_2}$  denotes the Euclidean norm of a vector. $ \otimes $  represents the Kronecker product. Given an $E \in {R^{m \times n}}$,${(E)^T} \in {R^{n \times m}}$ denotes its transpose. $\left[ {\rm E} \right]_i^{row}$ denotes the $i$-th row of ${\rm E}$ and ${[E]_{ij}}$ denotes the entry on $i$-th row and $j$-th column  of $E$. Let the distance from  $x \in {\Re ^N}$ to the set $C \subseteq {\Re ^N}$ be denoted by $dist(x,C) = \mathop {\inf }\limits_{y \in C} {\left\| {x - y} \right\|_2}$. Let the convex hull of a finite set of points $X = \left\{ {{x_1},...,{x_n}} \right\}$ be denoted by $Co(X) = \left\{ {\sum\nolimits_{i = 1}^n {{\alpha _i}{x_i}} \left| {{x_i} \in X,{\alpha _i} \in \Re ,} \right.{\alpha _i} \ge 0,\sum\nolimits_{i = 1}^n {{\alpha _i} = 1} } \right\}$. Finally, $\left| {\mathcal X} \right|$ denotes the cardinality of a set ${\mathcal X}$.

\section{Preliminaries} 
Let the communication topology among $n + m$ agents be represented by a fixed weighted directed graph $ \mathcal{G} = (\mathcal{V},\mathcal{E} ,\mathcal{A})$ with a set of nodes $\mathcal{V} = \left\{ {{\nu _1}, \ldots ,{\nu _{n + m}}} \right\}$, a set of edges $\mathcal{E}  \subseteq \mathcal{V} \times \mathcal{V}$, and a weighted adjacency matrix $\mathcal{A} = [{a_{ij}}]$ with non-negative adjacency elements ${a_{ij}}$, i.e., an edge $\left( {{\nu _j},{\nu _i}} \right) \in \mathcal{E} $  if and only if ${a_{ij}} > 0$ which means that the $i$-th agent can obtain information from the $j$-th agent, but not necessarily vice versa. We assume that there are no self-connections, i.e., $\left( {{\nu _i},{\nu _i}} \right) \notin \mathcal{E} $. If all of the diagonal elements of a square matrix are zero, the matrix is called a hollow matrix \cite{gentle2007matrix}. If no self-connections, the adjacency matrix $\mathcal{A}$, is a hollow matrix. Topological ordering of nodes in the graph $\mathcal{G}$ is an ordering labels of the nodes such that for every edge in $\mathcal{G}$, one has $\left( {{\nu _j},{\nu _i}} \right) \in \mathcal{E} :{\nu _j} < {\nu _i}$  \cite{goodrich2014data}. The set of neighbors of node ${\nu _i}$ is denoted by $N_i = \left\{ {{\nu _j} \in \mathcal{V}:\left( {{\nu _j},{\nu _i}} \right) \in \mathcal{E} ,j \ne i} \right\}$. The in-degree matrix of the graph $\mathcal{G}$  is defined as $\mathcal{D} = diag({d_i})$, where   ${d_i} = \sum\nolimits_{j \in N_i} {{a_{ij}}}$ is the weighted in-degree of node ${\nu _i}$.  A directed path $P_{ij}^\mathcal{G}({k_1},...,{k_\ell })$ from node ${\nu _i}$ to node  ${\nu _j}$ is a sequence of edges $\left( {{\nu _i},{\nu _{{k_1}}}} \right),\left( {{\nu _{{k_1}}},{\nu _{{k_2}}}} \right),...,\left( {{\nu _{{k_\ell }}},{\nu _j}} \right)$ with finite number of nodes ${\nu _{{k_m}}}$,  $m = 1,...,\ell $ in a directed graph $\mathcal{G}$. The length of a directed path $P_{ij}^\mathcal{G}({k_1},...,{k_\ell })$, denoted by $l(P_{ij}^\mathcal{G}({k_1},...,{k_\ell }))$, is the number of its nodes such that the repeated nodes count as one. The weight of a directed path $P_{ij}^\mathcal{G}({k_1},...,{k_\ell })$, denoted by $\mathcal{W}(P_{ij}^\mathcal{G}({k_1},...,{k_\ell }))$, is the product of the weights of its edges, i.e.,  $\mathcal{W}(P_{ij}^\mathcal{G}({k_1},...,{k_\ell })) = {a_{{k_1}i}}{a_{{k_2}{k_1}}}...{a_{j{k_\ell }}}$. If there exists at least one directed path from node ${\nu _i}$ to node ${\nu _j}$, then node ${\nu _j}$  is said to be reachable from node ${\nu _i}$. If node ${\nu _i}$ is the neighbor of node ${\nu _j}$, then node ${\nu _j}$ is said to be immediately reachable from node ${\nu _i}$. 


We assume that agents $1$ to $n$ are followers and form the set $\mathcal{F} := \left\{ {1,...,n} \right\}$, and agents $n + 1$  to $n + m$ are leaders and form the set $\mathcal{R} := \left\{ {n + 1,...,n + m} \right\}$. The set of the leaders that are neighbors of the $i$-th follower and the set of its influential followers are, respectively, denoted by $N_i^L = \left\{ {k \in \mathcal{R}:{a_{ik}} \ne 0} \right\}$ and $N_i^R = \left\{ {j \in \mathcal{F}:\mathcal{W}(P_{ij}^ \mathcal {G} ) \ne 0} \right\}$.
The Laplacian matrix $\mathcal{L} = [{\ell _{ij}}] \in {\Re ^{(n + m) \times (n + m)}}$ associated with $\mathcal{A}$ is defined as ${\ell _{ii}} = \sum\nolimits_{j \in N_i} {{a_{ij}}}$  and ${\ell _{ij}} =  - {a_{ij}}$ where $i \ne j$ and can be partitioned as 
\vspace{-0.2 cm}
\begin{align}
\mathcal{L} = \left[ {\begin{array}{*{20}{c}}
{{\mathcal{L}_1}}&{{\mathcal{L}_2}}\\
{{0_{m \times n}}}&{{0_{m \times m}}}
\end{array}} \right], \label{eq:1}
\end{align}
where ${\mathcal{L}_1} \in {\Re ^{n \times n}}$ and ${\mathcal{L}_2} \in {\Re ^{n \times m}}$. Similarly, the adjacency matrix of $\mathcal{A}$ and the in-degree matrix of  $\mathcal{D}$ associated with $\mathcal{G}$ can be partitioned as
\begin{align}
\mathcal {A} = \left[ {\begin{array}{*{20}{c}}
{\bar {\mathcal {A}}}&{ - {\mathcal {L}_2}}\\
{{0_{m \times n}}}&{{0_{m \times m}}}
\end{array}} \right],  
\quad {\mathcal {D}} = \left[ {\begin{array}{*{20}{c}}
{\bar {\mathcal {D}}}&{{0_{n \times m}}}\\
{{0_{m \times n}}}&{{0_{m \times m}}}
\end{array}} \right],  \label{eq:3}
\end{align}
where $\bar {\mathcal {A}} \in {\Re ^{n \times n}}$ captures the information flow among followers and $\bar {\mathcal {D}} \in {\Re ^{n \times n}}$.

In the sequel, we assume that the fixed communication graph $\mathcal{G}$ satisfies the following assumptions.
 
{\bf Assumption 1.} \label{as:1}
For each follower, there exists at least one leader that has a directed path to it.



{\bf Assumption 2.}\label{as:3}
Each agent has a unique label that is exchanged with its neighbors. Moreover, the type (leader or follower) of each agent is known to all the followers based on its label.

 { 
{\bf Remark 1.} In practice, a multi-agent system (e.g., smart grids, multi-robot systems, internet of things networks, etc) can be seen as a computerized system composed of multiple interacting cyber-physical agents that communicate information through a cyber layer. Therefore, it is reasonable to assume that each agent in the network has a unique address name like an IP address in computer networks (see \cite{pipattanasomporn2009multi,zhang2004multiagent}). Moreover, it is reasonable to assume that these address names convey some characteristic or identity-related information \cite{kendrick2017multi}, which can be decoded by a receiver agent. 
}

The following basic definitions and properties are taken from \cite{rosen2017handbook}. A partially ordered set (poset) $(\mathcal {X}, \le )$ is a pair consisting of a set $\mathcal {X}$, called the domain, and a partial ordering $\le$ on $\mathcal {X}$.  A totally ordered set is a poset in which every element is comparable to every other element. Every finite totally ordered set is called a well-ordered set. An order-preserving mapping from poset $(\mathcal {X}, \le )$ to poset $(\mathcal {Y}, \le )$ is a function $f:\mathcal {X} \to \mathcal {Y}$ such that ${x_i} \le {x_j}$ implies $f({x_i}) \le f({x_j})$. 

{\bf Definition 2.}\label{def:3}
Let $Q \subset \mathbb{N}$ be a non-empty finite set and $\mathcal {X} = \left\{ {{x_i} \in \mathbb{N} : {i \in Q,\,{x_i} \ne {x_j},\forall j \in Q\backslash \{ i\} }} \right\}$ be a finite subset of $\mathbb{N}$ with  $\left| Q \right|$ distinct elements. Define $\bar Q = \left\{ {1,...,\left| Q \right|} \right\}$. Then, the function $\vec X:\bar Q \to \mathcal {X}$ is called a sort function of the well-ordered set $(\mathcal {X}, \le )$ and is defined as $\vec {\mathcal {X}} := 
\{ (k,{x_k}): {x_k} \in \mathcal {X},{x_k} < {x_j},\,\forall \,k < j,\, k,j \in \bar Q\}.$

{\bf Lemma 1.}\label{lemma:1} \cite{meng2010distributed} 
Under Assumption 1, ${\mathcal {L}}_1$ is invertible, all the eigenvalues of ${\mathcal {L}}_1$ have positive real parts, each row of $- {\mathcal {L}}_1^{ - 1}{{\mathcal {L}}_2}$ has a sum equal to one, and each entry of $- {\mathcal {L}}_1^{ - 1}{{\mathcal {L}}_2}$ is nonnegative.

\section{Fully-Heterogeneous Containment Control Problem: A Novel Formulation} \label{Sec:2}
In this section, the fully-heterogeneous containment control problem (FHCCP) is formulated for linear multi-agent systems composed of $n$ non-identical followers and $m$ non-identical leaders.

Let the dynamics of the $i$-th follower be described by
\begin{align}
\left\{ \begin{array}{l}
{{\dot x}_i} = {A_i}{x_i} + {B_i}{u_i}\\
{y_i} = {C_i}{x_i},
\end{array} \right.,i \in \mathcal {F}, \label{eq:7}
\end{align}
where ${x_i} \in {\Re ^{{N_i}}}$ is the state of the $i$-th agent, ${u_i} \in {\Re ^{{P_i}}}$ is its input, and ${y_i} \in {\Re ^Q}$ is its output. The leaders' trajectories are assumed to be generated by the following exo-system dynamics.
\begin{align}
\left\{ {\begin{array}{*{20}{c}}
{{{\dot \omega }_k} = {S_k}{\omega _k}}\\
{{y_k} = {D_k}{\omega _k}}
\end{array}} \right.,k \in \mathcal {R}, \label{eq:8}
\end{align}
where ${\omega _k} \in {\Re ^q}$ and ${y_k} \in {\Re ^Q}$ are, respectively, the state and output of the $k$-th agent.


{\bf Assumption 3.}\label{as:4}
  $({A_i},{B_i})$ in (\ref{eq:7}) is stabilizable $\forall i \in \mathcal {F}$. Moreover, ${C_i}$ in  (\ref{eq:7}) is full row rank $\forall i \in \mathcal {F}$.


{\bf Assumption 4.}\label{as:6}
The leaders' dynamics are marginally stable.

{\bf Problem 1.}\label{prb:1}
Consider the multi-agent system (\ref{eq:7})-(\ref{eq:8}). Design the control protocols ${u_i}$, $\forall i \in \mathcal {F}$ so that the outputs of the followers converge to the convex hull spanned by outputs of the leaders. That is,
\begin{align}
\mathop {lim}\limits_{t \to \infty } \,dist({y_i}(t),Co({y_k}(t),k \in \mathcal {R})) = 0,\,\,\,\,\forall i \in \mathcal {F}. \label{eq:9}
\end{align} 

To solve this problem, we first define a new local containment error for the  $i$-th follower as
\begin{align}
&{\bar e_i}(t) = {y_i}(t) - y_i^*(t), \label{eq:10} \\
& y_i^* = ([{\Phi _P}]_i^{row} \otimes {I_Q}){{D^*}\bar \omega} \label{eq:21} \\
&[{\Phi _P}]_i^{row} = [{- {\mathcal {L}}_1^{ - 1}{{\mathcal {L}}_2}}]_i^{row}  \label{eq:38}
\end{align}
 with ${{\mathcal {L}}_1}$ and ${{\mathcal {L}}_2}$ defined in (\ref{eq:1}), ${D^*} = diag({D_{n + 1}},...,\allowbreak{D_{n + m}})$, and $\bar \omega  = col({\omega _{n + 1}},...,{\omega _{n + m}})$.



{\bf Lemma 2.}\label{cor:1}
Under Assumption 1, let
\begin{align}
\mathop {lim}\limits_{t \to \infty } \,{\bar e_i}(t) = 0,\forall i \in \mathcal {F}, \label{eq:19*}
\end{align}
where ${\bar e_i}(t)$ is defined in (\ref{eq:10}). Then, Problem 1 is solved.

\textbf{Proof.} The proof follows from (\ref{eq:10}), the definition of convex hull, and the results of Lemma 1.\hfill {$\square$}


{\bf Problem 2.} \label{prb:2}(Multiple reference trajectories tracking problem for FHCCP)
Consider the MAS (\ref{eq:7})-(\ref{eq:8}). Design the control protocols ${u_i}$, $\forall i \in \mathcal {F}$, so that (\ref{eq:19*}) is satisfied.

Using (\ref{eq:8}) and (\ref{eq:21}), the virtual exo-system dynamics that generates the reference trajectory  $y_i^*$ (\ref{eq:21}) for the $i$-th follower is given by
\begin{align}
\left\{ \begin{array}{l}
\dot {\bar \omega}  = {S^*}\bar \omega \\
y_i^* = ([{\Phi _P}]_i^{row} \otimes {I_Q}){D^*}\bar \omega 
\end{array} \right.,\forall i \in \mathcal {F} \label{eq:22}
\end{align}
where ${S^*} = diag({S_{n + 1}},...,{S_{n + m}})$, ${D^*} = diag({D_{n + 1}},...,\allowbreak{D_{n + m}})$, and $\bar \omega  = col({\omega _{n + 1}},...,{\omega _{n + m}})$. 


{
{\bf Definition 3.}\label{def:4}
The normalized level of influences (NLIs) of all leaders on the $i$-th follower is defined by 
\begin{align}
[{\Phi _P}]_i^{row} =& [{- {\mathcal {L}}_1^{ - 1}{{\mathcal {L}}_2}}]_i^{row} \nonumber \\
=&[{\varphi _{i1}}, ..., {\varphi _{i(k - n)}},...,{\varphi _{im}}] \in {\Re ^{1 \times m}}  , k \in \mathcal {R}, \label{eq:20}
\end{align}
 where ${\varphi _{i(k - n)}} \in \Re$ is the normalized level of influence (NLI) of the $k$-th leader on the $i$-th follower, $m = \left| \mathcal {R} \right|$, and $n = \left| \mathcal {F} \right|$.}


\section{A Distributed algorithm for calculating the NLIs of the influential leaders on each follower} \label{Sec:4}


 

{To find the NLIs of the influential leaders on the $i$-th follower, the notion of the local graph from the $i$-th follower perspective is defined as follows.}

 {{\bf Definition 4.}\label{def:5}
The local graph from the $i$-th follower perspective, denoted by ${\bar {\mathcal {G}}_i} = ({\bar {\mathcal{V}}_i},{\bar {\mathcal{E}} _i},{\mathcal {A}}_\ell ^i)$, is a sub-graph of overall graph $\mathcal{G}$ and it is obtained by removing non-influential nodes of the $i$-th follower along with their corresponding edges from $\mathcal{G}$. }


In order to obtain the Laplacian matrix of graph ${\bar {\mathcal {G}}_i}$ which leads to obtaining the NLIs of leaders on the $i$-th follower, agent $i$ needs to find out about the set of the labels of the influential leaders and followers, i.e.,  ${\bar {\mathcal{V}}_i}$, and the set of their corresponding edges, i.e., ${\bar {\mathcal{E}} _i}$. To this end, the following definition plays a critical role.

{\bf Definition 5.}\label{def:6}
The set of labels of the influential leaders and followers of the $i$-th follower is defined as ${\bar {\mathcal{V}}_i} = \bar N_i^A \cup \bar N_i^L$
where $\bar N_i^A: = \mathop  \cup \limits_{t \in N_i^R \cup \left\{ i \right\}} N_t^A$
is the set of labels of the influential followers of the $i$-th follower, including itself, where $N_i^A: = \left\{ {\left\{ i \right\} \cup N_i} \right\}\backslash N_i^L$
is the set of labels of followers agents that the $i$-th follower is immediately reachable from, including itself, and $\bar N_i^L: = \mathop  \cup \limits_{t \in N_i^R \cup \{ i\} } N_t^L$
is the set of labels of all the influential leaders of the $i$-th follower.

Note that to find $\bar N_i^A$ and $\bar N_i^L$, one should have the whole graph topology of ${\bar {\mathcal {G}}_i}$. Algorithm 1 is utilized for constructing these sets in a fully distributed manner.

\small{
\begin{algorithm}[!ht]
\caption{\small{Distributed algorithm for constructing $\bar N_i^A$ and $\bar N_i^L$, $\forall i$.} }
\begin{enumerate}
\item [1:] Initialize $\bar N{_i^{L^{(0)}}} = N_i^L$ and $\bar N{_i^{A^{(0)}}} = N_i^A$.
\item [2:] \textbf{procedure} {$\forall i = 1,...,n$}{}
\item [3:] Exchange information with neighbors to compute
 		\begin{align}
 		\bar N{_i^{L^{(k)}}}: = \mathop  \cup \limits_{j \in N_i\backslash N_i^L} \bar N{_j^{L^{(k - 1)}}} \cup \bar N{_i^{L^{(0)}}}, \label{eq:60} \\
 		\bar N{_i^{A^{(k)}}}: = \mathop  \cup \limits_{j \in N_i\backslash N_i^L} \bar N{_j^{A^{(k - 1)}}} \cup \bar N{_i^{A^{(0)}}}.\label{eq:61}
 		\end{align}.
\item [4:] Let $k: = k + 1$ and repeat step 3 until $\bar N{_i^{L^{(k)}}} - \bar N{_i^{L^{(k - 1)}}} = \emptyset$, $\bar N{_j^{L^{(k - 1)}}} - \bar N{_j^{L^{(k - 2)}}} = \emptyset $, $\bar N{_i^{A^{(k)}}} - \bar N{_i^{A^{(k - 1)}}} = \emptyset $, and $\bar N{_j^{A^{(k - 1)}}} - \bar N{_j^{A^{(k - 2)}}} = \emptyset$, $\forall j \in N_i\backslash N_i^L$.
\item [5:] On convergence set $\bar N_i^L = \bar N{_i^{L^{(k)}}}$ and $\bar N_i^A = \bar N{_i^{A^{(k)}}}$.
\item [6:] \textbf{end procedure}
\end{enumerate}
\end{algorithm}
}




{\bf Definition 6.}\label{def:7}
Let function $\bar {\mathcal{E}} _i^{\mathcal {A}}:h_i^{\mathcal {A}} \to {\mathcal{V}}_i^{\mathcal {A}}$ be defined as $\bar {\mathcal{E}} _i^{\mathcal {A}}: = \left\{ {((i,j),{a_{ij}}):j \in N_i} \right\}$
with the domain of $h_i^{\mathcal {A}}: = \left\{ {(i,j):j \in N_i} \right\}$
and the range of ${\mathcal{V}}_i^{\mathcal {A}}: = \left\{ {({a_{ij}}):j \in N_i} \right\}.$
Then, the function ${\bar {\mathcal{E}} _i}:{h_i} \to {{\mathcal{V}}_i}$ is defined as ${\bar {\mathcal{E}} _i}: = \mathop  \cup \limits_{t \in N_i^R \cup \left\{ i \right\}} \bar {\mathcal{E}} _t^{\mathcal {A}}$
with the domain and range of ${h_i}: = \mathop  \cup \limits_{t \in N_i^R \cup \left\{ i \right\}} h_i^{\mathcal {A}}$ and ${{\mathcal{V}}_i}: = \mathop  \cup \limits_{t \in N_i^R \cup \left\{ i \right\}} {\mathcal{V}}_i^{\mathcal {A}}$, respectively.

Note that to find ${\bar {\mathcal{E}} _i}$, one should have the whole graph topology of ${\bar {\mathcal {G}}_i}$. Algorithm 2 is utilized for constructing the set of ${\bar {\mathcal{E}} _i}$, $\forall i = 1,...,n$ in a fully distributed manner.

\small{
\begin{algorithm}[!ht]
\caption{\small {Distributed algorithm for constructing ${\bar {\mathcal{E}} _i}$, $\forall i$.}}
\begin{enumerate}
\item [1:] Initialize ${\bar {\mathcal{E}} _i}^{(0)} = \bar {\mathcal{E}} _i^{\mathcal {A}}$.          
\item [2:] \textbf{procedure} {$\forall i = 1,...,n$}{}
\item [3:] Exchange information with neighbors to compute
		\begin{align}
		{\bar {\mathcal{E}} _i}^{(k)}: = \mathop  \cup \limits_{j \in N_i\backslash N_i^L} {\bar {\mathcal{E}} _j}^{(k - 1)} \cup {\bar {\mathcal{E}} _i}^{(0)}. \label{eq:67}
		\end{align}
\item [4:] Let $k: = k + 1$ and repeat step 3 until ${\bar {\mathcal{E}} _i}^{(k)} - {\bar {\mathcal{E}} _i}^{(k - 1)} = \emptyset $, ${\bar {\mathcal{E}} _j}^{(k - 1)} - {\bar {\mathcal{E}} _j}^{(k - 2)} = \emptyset$, $\forall j \in N_i\backslash N_i^L$.
\item [5:] On convergence set ${\bar {\mathcal{E}} _i} = {\bar {\mathcal{E}} _i}^{(k)}$.
\item [6:] \textbf{end procedure}
\end{enumerate}
\end{algorithm}
}

{\bf Proposition 1.}\label{theorem:8*}
Let $\mathcal{G}$ be the communication graph with a fixed topology and finite number of nodes. Then, under Assumptions~1-2, $\forall i = 1,...,n$, (a) $\bar N{_i^{A^{(k)}}} \to \bar N_i^A$, (b) $\bar N{_i^{L^{(k)}}} \to \bar N_i^L$, and (c) ${\bar {\mathcal{E}} _i}^{(k)} \to {\bar {\mathcal{E}} _i}$ after a finite number of iterations which is bounded by the length of the longest path in the communication graph ${\bar {{\mathcal {G}}}_i}$. 
	
	
\textbf{Proof.}
The proof is easy to be obtained and omitted. \hfill {$\square$}

{\bf Remark 2.}\label{rem:5}
 {Note  that the $i$-th follower agent does not need to exchange the information of $N_i^L$, $N_i^A$, $\bar N_j^L$, $\bar N_j^A$, $\bar {\mathcal{E}} _i^A$, and ${\bar {\mathcal{E}} _j}$, $\forall j \in N_i\backslash N_i^L$ all the time with its neighbors, and after some iterations (depends on the longest path in communication graph ${\bar {\cal G}_i}$), once it identifies its ${\bar {\cal G}_i}$, it can stop exchanging this information which does not significantly increase the communication burden.}





Suppose that after a sufficient number of iterations, $\bar N{_i^{A^{(k)}}} \to \bar N_i^A$, $\bar N{_i^{L^{(k)}}} \to \bar N_i^L$, and ${\bar {\mathcal{E}} _i}^{(k)} \to {\bar {\mathcal{E}} _i}$, $\forall i = 1,...,n$ in Algorithms 1 and 2. To obtain the Laplacian matrix of graph ${\bar {\mathcal {G}}_i}$  using these sets, which leads to obtaining the NLIs of leaders on the $i$-th follower, one first needs to find the weighted adjacency matrix and local in-degree matrix of ${\bar {\mathcal {G}}_i}$, i.e.,  $\mathcal {A}_\ell ^i$ and $\mathcal {D}_\ell ^i$, which are defined as follows.

{\bf Definition 7.}\label{def:8}
The local weighted adjacency matrix of graph $\mathcal{G}$ from the $i$-th follower's perspective is defined as
\begin{align}
{\mathcal {A}}_\ell ^i = \left[ {\begin{array}{*{20}{c}}
{\bar {\mathcal {A}}_\ell ^i}&{\bar {\mathcal {A}}_{2\ell }^i}\\
{{0_{{{\bar {\bar l}}_i} \times {l_i}}}}&{{0_{{{\bar {\bar l}}_i} \times {{\bar {\bar l}}_i}}}}
\end{array}} \right] \in {\Re ^{({l_i} + {{\bar {\bar l}}_i}) \times ({l_i} + {{\bar {\bar l}}_i})}} \label{eq:68}
\end{align}
where
\begin{align}
\bar {\mathcal {A}}_\ell ^i = \left[ {\begin{array}{*{20}{c}}
{{a_{{\mu _i}(1){\mu _i}(1)}}}& \cdots &{{a_{{\mu _i}(1){\mu _i}({l_i})}}}\\
 \vdots & \ddots & \vdots \\
{{a_{{\mu _i}({l_i}){\mu _i}(1)}}}& \cdots &{{a_{{\mu _i}({l_i}){\mu _i}({l_i})}}}
\end{array}} \right] \in {\Re ^{{l_i} \times {l_i}}} \label{eq:69}
\end{align}
with
\begin{align}
{a_{{\mu _i}(k){\mu _i}(m)}}: = \left\{ {\begin{array}{*{20}{c}}
{{{\bar {\mathcal{E}} }_i}({\mu _i}(k),{\mu _i}(m))}&{({\mu _i}(k),{\mu _i}(m)) \in {h_i}}\\
0&{({\mu _i}(k),{\mu _i}(m)) \notin {h_i}}
\end{array}} \right. \label{eq:69*}
\end{align}
where $k = 1,...,{l_i},m = 1,...,{l_i}$, is the weighted adjacency sub-matrix of the graph ${\bar {\mathcal {G}}_i}$ relating all the influential followers of the $i$-th follower, including itself, and ${\mu _i}: = \vec {\bar N}_i^A$
is the sort function of well-ordered set $(\bar N_i^A, \le )$, and ${l_i} = \left| {\bar N_i^A} \right|$ is the number of influential followers of the  $i$-th follower agent plus itself. Moreover,
\begin{align}
\bar {\mathcal {A}}_{2\ell }^i = \left[ {\begin{array}{*{20}{c}}
{{a_{{\mu _i}(1)\bar \mu _i^{\cal T}(1)}}}& \cdots &{{a_{{\mu _i}(1)\bar \mu _i^{\cal T}({{\bar {\bar l}}_i})}}}\\
 \vdots & \ddots & \vdots \\
{{a_{{\mu _i}({l_i})\bar \mu _i^{\cal T}(1)}}}& \cdots &{{a_{{\mu _i}({l_i})\bar \mu _i^{\cal T}({{\bar {\bar l}}_i})}}}
\end{array}} \right] \in {\Re ^{{l_i} \times {{\bar {\bar l}}_i}}} \label{eq:71}
\end{align}
with 
\begin{align}
{a_{{\mu _i}(k)\bar \mu _i^{\cal T}(m)}}: = \left\{ {\begin{array}{*{20}{c}}
{{{\bar {\mathcal{E}} }_i}({\mu _i}(k),\bar \mu _i^{\cal T}(m))}&{({\mu _i}(k),\bar \mu _i^{\cal T}(m)) \in {h_i}}\\
0&{({\mu _i}(k),\bar \mu _i^{\cal T}(m)) \notin {h_i}}
\end{array}} \right. \label{eq:71*}
\end{align}
where $k = 1,...,{l_i},m = 1,...,{\bar {\bar l}_i}$, is the weighted adjacency sub-matrix of the graph ${\bar {\mathcal {G}}_i}$ relating all the influential leaders of the $i$-th follower, where $\bar \mu _i^{\cal T}: = \vec {\bar N}_i^L$
is the sort function of the well-ordered set $(\bar N_i^L, \le )$, and ${\bar {\bar l}_i} = \left| {\bar N_i^L} \right|$ is the number of influential leaders of the $i$-th follower.

{\bf Definition 8.}\label{def:9}
The local in-degree matrix of the graph $\mathcal{G}$ from the $i$-th follower's perspective is defined as
\begin{align}
{\mathcal {D}}_\ell ^i = \left[ {\begin{array}{*{20}{c}}
{\bar {\mathcal {D}}_\ell ^i}&{{0_{{l_i} \times {{\bar {\bar l}}_i}}}}\\
{{0_{{{\bar {\bar l}}_i} \times {l_i}}}}&{{0_{{{\bar {\bar l}}_i} \times {{\bar {\bar l}}_i}}}}
\end{array}} \right] \in {\Re ^{({l_i} + {{\bar {\bar l}}_i}) \times ({l_i} + {{\bar {\bar l}}_i})}} \label{eq:73}
\end{align}
where 
\vspace{-0.05in}
\begin{align}
\bar {\mathcal {D}}_\ell ^i = \left[ {\begin{array}{*{20}{c}}
{\bar d_1^i}&0&0\\
0& \ddots &0\\
0&0&{\bar d_{{l_i}}^i}
\end{array}} \right] \in {\Re ^{{l_i} \times {l_i}}} \label{eq:74}
\end{align}
with
\begin{align}
\bar d_k^i = \sum\nolimits_{m = 1}^{{l_i}} {{a_{{\mu _i}(k){\mu _i}(m)}}}  + \sum\nolimits_{m = 1}^{{{\bar {\bar l}}_i}} {{a_{{\mu _i}(k)\bar \mu _i^{\cal T}(m)}}} , \label{eq:75}
\end{align}
$\forall k \in \{ 1,...,{l_i}\}$ is the in-degree sub-matrix of the graph ${\bar {\mathcal {G}}_i}$.


{\bf Definition 9.}\label{def:10}
The local Laplacian matrix of the graph $\mathcal{G}$ from the $i$-th follower's perspective is defined as
\begin{align}
\bar {\mathcal {L}}_\ell ^i = \left[ {\begin{array}{*{20}{c}}
{\bar {\mathcal {L}}_{1\ell }^i}&{\bar {\mathcal {L}}_{2\ell }^i}\\
{{0_{{{\bar {\bar l}}_i} \times {l_i}}}}&{{0_{{{\bar {\bar l}}_i} \times {{\bar {\bar l}}_i}}}}
\end{array}} \right] \in {\Re ^{({l_i} + {{\bar {\bar l}}_i}) \times ({l_i} + {{\bar {\bar l}}_i})}} \label{eq:76}
\end{align}
where 
\begin{align}
\bar {\mathcal {L}}_{1\ell }^i &= \bar {\mathcal {D}}_\ell ^i - \bar {\mathcal {A}}_\ell ^i \in {\Re ^{{l_i} \times {l_i}}}, \label{eq:77} \\
\bar {\mathcal {L}}_{2\ell }^i &=  - \bar {\mathcal {A}}_{2\ell }^i \in {\Re ^{{l_i} \times {{\bar {\bar l}}_i}}}, \label{eq:78}
\end{align}
are Laplacian sub-matrices of the graph ${\bar {\mathcal {G}}_i}$. 


{\bf Definition 10.}\label{def:11}
The NLIs of influential leaders of the $i$-th follower, denoted by $\Phi _{P\ell }^i$ is a row vector that represents the NLIs of leaders on $i$-th follower based on ${\bar {\mathcal {G}}_i}$, and is defined as 
\begin{align}
\Phi _{P\ell }^i = [{\varphi _{i\bar \mu _i^{\cal T}(1)}},...,{\varphi _{i\bar \mu _i^{\cal T}({{\bar {\bar l}}_i})}}] \in {\Re ^{1 \times {{\bar {\bar l}}_i}}} \label{eq:79}
\end{align}
where ${\varphi _{i\bar \mu _i^{\cal T}(m)}}$, $\forall m = 1,...,{\bar {\bar l}_i}$ is the NLI of the $\bar \mu _i^{\cal T}(m)$-th leader on  the $i$-th follower. 

	
{\bf Lemma 3.}\label{lemma:10*}
Consider the elements of the local Laplacian matrix (\ref{eq:76}) for agent  $i$. Then, the NLI of its influential leaders, defined in (\ref{eq:79}), is obtained as
\begin{align}
\Phi _{P\ell }^i = - {\Upsilon _i}\bar {\mathcal {L}}{{_{1\ell }^i}^{ - 1}}\bar {\mathcal {L}}_{2\ell }^i  \label{eq:79*}
\end{align}
where ${\Upsilon _i} \in {\Re ^{1 \times {l_i}}}$ is a row vector, with elements of 
\begin{align}
{\left[ {{\Upsilon _i}} \right]_{1m}} = \left\{ {\begin{array}{*{20}{c}}
{1\,\,\,\,,}&{{\mu _i}(m) = i}\\
{0\,\,\,,}&{Otherwise}
\end{array}} \right.,\forall m = 1,...,{l_i}. \label{eq:84}
\end{align}

\textbf{Proof.} Using (\ref{eq:38}), (\ref{eq:76}), and Definitions~4 and 10, one can observe that $\Phi _{P\ell }^i = [ { - \bar {\mathcal {L}}{{_{1\ell }^i}^{ - 1}}\bar {\mathcal {L}}_{2\ell }^i} ]_m^{row}$, where ${\mu _i}(m) = i$, $\forall m = 1,...,{l_i}$. Note that ${\Upsilon _i}$ is a row vector, for which all of its elements are zero except the one that is related to the $i$-th follower. This completes the proof. \hfill {$\square$}

Algorithm 3 presents a distributed algorithm for calculating (\ref{eq:77}), (\ref{eq:78}), and  (\ref{eq:79*}) by the $i$-th follower, $\forall i = 1,...,n$ in a distributed fashion. 

\small{
\begin{algorithm}[!ht]
\caption{ \small {Distributed algorithm for constructing $\bar {\mathcal {L}}_{1\ell }^i$, $\bar {\mathcal {L}}_{2\ell }^i$, and $\Phi _{P\ell }^i$.}}
\begin{enumerate}
\item [1:] Calculate $\bar N_i^A$, $\bar N_i^L$, and ${\bar {\mathcal{E}} _i}$ as Algorithms 1 and 2, respectively.
\item [2:] Construct $\bar {\mathcal {A}}_\ell ^i$ and  $\bar {\mathcal {A}}_{2\ell }^i$  as follows: 
\begin{enumerate}
				\item Set $\bar {\mathcal {A}}_\ell ^i = {0_{{l_i} \times {l_i}}}$ and $\bar {\mathcal {A}}_{2\ell }^i = {0_{{l_i} \times {{\bar {\bar l}}_i}}}$.               
				\item Change the elements of $\bar {\mathcal {A}}_\ell ^i$ and $\bar {\mathcal {A}}_{2\ell }^i$ as (\ref{eq:69*}) and (\ref{eq:71*}), respectively.
				\end{enumerate}         
\item [3:] Construct $\bar {\mathcal {D}}_\ell ^i$ as follows:
				\begin{enumerate}
				\item Set $\bar {\mathcal {D}}_\ell ^i = {0_{{l_i} \times {l_i}}}$.
				\item Calculate $\bar d_k^i$, $\forall k \in \{ 1,...,{l_i}\} $ as (\ref{eq:75}).
				\item Change the elements of $\bar {\mathcal {D}}_\ell ^i$ as (\ref{eq:74}).
				\end{enumerate}
\item [4:] Construct $\bar {\mathcal {L}}_{1\ell }^i$ and  $\bar {\mathcal {L}}_{2\ell }^i$ as follows: 
				\begin{enumerate}
		    	\item Calculate $\bar {\mathcal {L}}_{1\ell }^i$  and  $\bar {\mathcal {L}}_{2\ell }^i$  as (\ref{eq:77}) and (\ref{eq:78}), respectively.
				\item Set ${\Upsilon _i} = {0_{1 \times {l_i}}}$.
				\item Change the elements of  ${\Upsilon _i}$ as (\ref{eq:84}).
				\end{enumerate} 
\item [5:] Calculate $\Phi _{P\ell }^i$ as (\ref{eq:79*}).
\end{enumerate}
\end{algorithm}
}

The virtual exo-system dynamics for the $i$-th follower in (\ref{eq:22}) includes the dynamics of all leaders, even if they do not have any influence on its output. It follows from Definitions~4 and 10 that this virtual exo-system dynamics can be reduced to only include the dynamics of the influential leaders of the $i$-th follower, i.e., $\bar N_i^L$, since the non-influencing leaders have no effect on its reference trajectory $y_i^*$ in (\ref{eq:1}). To this end, the corresponding dynamics of the leaders with zero influence on the $i$-th follower and their states can be removed from (\ref{eq:22}), without changing its reference trajectory, resulting in the following equivalent virtual exo-system
\begin{align}
\left\{ \begin{array}{l}
\dot \Omega _i^R = \bar {\bar S}_i^R\Omega _i^R\\
y_i^* = (\Phi _{P\ell }^i \otimes {I_Q})\bar {\bar D}_i^R\Omega _i^R
\end{array} \right. \label{eq:85}
\end{align}
where $\Omega _i^R = col({\omega _{\bar \mu _i^{\cal T}(1)}},...,{\omega _{\bar \mu _i^{\cal T}({{\bar {\bar l}}_i})}})$, $\bar {\bar S}_i^R = diag({S_{\bar \mu _i^{\cal T}(1)}},...\allowbreak,{S_{\bar \mu _i^{\cal T}({{\bar {\bar l}}_i})}})$ and $\bar {\bar D}_i^R = diag({D_{\bar \mu _i^{\cal T}(1)}},...,{D_{\bar \mu _i^{\cal T}({{\bar {\bar l}}_i})}})$ are the augmented state and augmented dynamics of the influential leaders of the $i$-th follower.

Using Definition~10, the local containment error (\ref{eq:10}) can now be rewritten as
\vspace{-0.1in}
\begin{align}
{\bar e_i} = {y_i} - (\Phi _{P\ell }^i \otimes {I_Q})\bar {\bar D}_i^R\Omega _i^R. \label{eq:86}
\end{align}

The $i$-th follower can build the virtual exo-system (\ref{eq:85}), only if Assumption 1  holds. The following lemma  gives the $i$-th follower an approach to test the validity of Assumption 1 in a distributed manner.

{\bf Lemma 4.}\label{lemma:10}
Under Assumption 2, $\Phi _{P\ell }^i{1_{{{\bar {\bar l}}_i}}} = 1, \forall i \in \mathcal {F}$,
if and only if Assumption~1 holds.

\textbf{Proof.} Sufficiency. When Assumption~1 holds for the graph $\mathcal{G}$, based on Definition~4 there exists at least one leader that it influences the $i$-th follower in the sub-graph ${\bar {\mathcal{G}}_i}$. Therefore, Assumption~1 is also satisfied for the sub-graph ${\bar {\mathcal{G}}_i}$, $\forall i \in {\mathcal{F}}$. Thus, using the Theorem 3.1 of \cite{cao2012distributed} for ${\bar {\mathcal{G}}_i}$, each entry of ${\bar {\mathcal{L}}{_{1\ell }^i}^{ - 1}}\bar {\mathcal{L}}_{2\ell }^i \in {\Re ^{{l_i} \times {{\bar {\bar l}}_i}}}$, $\forall i \in {\mathcal{F}}$  is non-negative, and each row of it has a sum equal to one, which results in $\Phi _{P\ell }^i{1_{{{\bar {\bar l}}_i}}} = 1, \forall i \in \mathcal {F}$.

Necessity. Assume that $\Phi _{P\ell }^i{1_{{{\bar {\bar l}}_i}}} = 0$. Based on Definitions~4 and 10, it implies that the $i$-th follower is not under the influence of any leader in the graph ${\bar {\mathcal{G}}_i}$ and consequently in the graph $\mathcal{G}$. This completes the proof. \hfill {$\square$}


 {
  Although, the highest required number of iterations to $\bar N_i^{{A^{(k)}}} \to \bar N_i^A$,  $\bar N_i^{{L^{(k)}}} \to \bar N_i^L$, and ${\overline {\cal E} _i}^{(k)} \to {\overline {\cal E} _i}$  is bounded by the longest path (based on Proposition 1), the presented method does not search for finding the longest path, and we just use the union operations to collect information from the influential nodes to construct the required sets in Algorithms 1 and 2. Moreover, note that the length of a path between two nodes can be seen as the number of the non-repeated distinctive nodes in it. This implies that the length of a path even with cycles is finite and is equal to the finite number of the distinctive nodes in it which are the subset of the set of nodes ${\cal V}$. We now analyze the scalability of Algorithms 1 and 2 to measures the growth of its complexity (i.e., the number of iterations) in relation to the growth of the size of the graph and show that Algorithms 1 and 2 roughly uses $O(n)$ time complexity and therefore these algorithms are quite efficient in terms of  scalability. To this aim, let the scalability of Algorithms 1 and 2 for finding mentioned sets in the graph ${\cal G}$  be defined as ${\cal S}({\cal G}) = {{{\cal T}({\cal G})} \mathord{\left/
 {\vphantom {{{\cal T}({\cal G})} {{\rm{size}}({\cal G})}}} \right.
 \kern-\nulldelimiterspace} {{\rm{size}}({\cal G})}}$
 where ${\cal T}({\cal G})$  is the highest required number of iterations to $\bar N_i^{{A^{(k)}}} \to \bar N_i^A$,  $\bar N_i^{{L^{(k)}}} \to \bar N_i^L$, and ${\overline {\cal E} _i}^{(k)} \to {\overline {\cal E} _i}$, $\forall i \in {\cal F}$. 
Defining the size of a graph as $\left| {\cal V} \right| + \left| {\cal E} \right|$, we now summarize the instanced-based scalability ${\cal S}({\cal G})$ over all different size of graphs as ${\cal S}(n) = {\sup _{\forall {\cal G} \in {{\cal G}_n}}}{{{\cal T}({\cal G})} \mathord{\left/
 {\vphantom {{{\cal T}({\cal G})} {{\rm{size}}({\cal G})}}} \right.
 \kern-\nulldelimiterspace} {{\rm{size}}({\cal G})}}$
where ${G_n}$ denotes the set of all directed graph with size $n$, and ${\sup _{\forall {\cal G} \in {{\cal G}_n}}}{\cal T}({\cal G})$  denote the (worst-case) highest required number of iterations to $\bar N_i^{{A^{(k)}}} \to \bar N_i^A$,  $\bar N_i^{{L^{(k)}}} \to \bar N_i^L$, and ${\overline {\cal E} _i}^{(k)} \to {\overline {\cal E} _i}$,  $\forall i \in {\cal F}$ on the set of all directed graph with size $n$. Therefore, ${\cal S}(n) = {{{\cal T}(n)} \mathord{\left/
 {\vphantom {{{\cal T}(n)} n}} \right.
 \kern-\nulldelimiterspace} n}$. 
Based on Proposition 1 and the definition of the length of a path, one has ${\cal T}(n) \le {\sup _{\forall i \in {\cal F}}}{\mkern 1mu} {\mkern 1mu} l_i^{\overline {\cal G} _i^*} \le n$  for a graph  ${\cal G}$ with the size of $n$.  This implies that for general directed graph one can implement Algorithms 1 and 2 which roughly uses $O(n)$ time complexity (i.e.,polynomial time iterations), and therefore this approach is quite efficient in terms of scalability.}

\section{Adaptive distributed observer of virtual exo-system trajectories: Heterogeneous leaders case}\label{Sec:5}

To estimate the virtual exo-system trajectories and dynamics for each follower, consider the following adaptive distributed observer
\begin{align}
{\dot \eta _i} &= {\bar S_i}{\eta _i} + \beta _{}^\eta (\sum\limits_{j \in N_i\backslash N_i^L} {{a_{ij}}} (\eta _j^i - {\eta _i}) + {\delta _i}(\Omega _i^R - {\eta _i})), \label{eq:92} \\
{\dot S_i} &= \beta _{}^S(\sum\limits_{j \in N_i\backslash N_i^L} {{a_{ij}}} (S_j^i - {S_i}) + {\delta _i}(\bar S_i^R - {S_i})), \label{eq:93} \\
{\dot D_i} &= \beta _{}^D(\sum\limits_{j \in N_i\backslash N_i^L} {{a_{ij}}} (D_j^i - {D_i}) + {\delta _i}(\bar D_i^R - {D_i})), \label{eq:94}
\end{align}
where $\beta _{}^\eta  > 0$, $\beta _{}^S > 0$, $\beta _{}^D > 0$, ${\delta _i} = diag({a_{i\bar \mu _i^{\cal T}(1)}},...,{a_{i\bar \mu _i^{\cal T}({{\bar {\bar l}}_i})}})$ is diagonal matrix with diagonal entries of the pining gains of the leaders that the $i$-th follower is immediately reachable from, ${\eta _i}$ is the stack column of $\eta _i^{\bar \mu _i^{\cal T}({k_i})}$, for ${k_i} = 1,...,{\bar {\bar l}_i}$, i.e., ${\eta _i} = col(\eta _i^{\bar \mu _i^{\cal T}(1)}, \ldots ,\eta _i^{\bar \mu _i^{\cal T}({{\bar {\bar l}_i}})})$, where $\eta _i^{\bar \mu _i^{\cal T}({k_i})}$ is the state estimation of the $\bar \mu _i^{\cal T}({k_i})$-th leader by the $i$-th follower. Moreover, ${S_i}$ and ${D_i}$ are the stack columns of $S_i^{\bar \mu _i^{\cal T}({k_i})}$ and $D_i^{\bar \mu _i^{\cal T}({k_i})}$ for ${k_i} = 1,...,{\bar {\bar l}_i}$, i.e., ${S_i} = col(S_i^{\bar \mu _i^{\cal T}(1)}, \ldots ,S_i^{\bar \mu _i^{\cal T}({{\bar {\bar l}}_i})})$  and ${D_i} = col(D_i^{\bar \mu _i^{\cal T}(1)}, \ldots ,D_i^{\bar \mu _i^{\cal T}({{\bar {\bar l}}_i})})$, where $S_i^{\bar \mu _i^{\cal T}({k_i})}$  and $D_i^{\bar \mu _i^{\cal T}({k_i})}$  are the dynamics estimation of the $\bar \mu _i^{\cal T}({k_i})$-th leader by the $i$-th follower, respectively, and ${\bar S_i} = diag(S_i^{\bar \mu _i^{\cal T}(1)}, \ldots ,S_i^{\bar \mu _i^{\cal T}({{\bar {\bar l}}_i})})$  and ${\bar D_i} = diag(D_i^{\bar \mu _i^{\cal T}(1)}, \ldots ,D_i^{\bar \mu _i^{\cal T}({{\bar {\bar l}}_i})})$. Furthermore, $\Omega _i^R$ is the stack column of ${\omega _{\bar \mu _i^{\cal T}(j)}}$, for ${k_i} = 1,...,{\bar {\bar l}_i}$, i.e., $\Omega _i^R = col({\omega _{\bar \mu _i^{\cal T}(1)}},...,{\omega _{\bar \mu _i^{\cal T}({{\bar {\bar l}}_i})}})$, where ${\omega _{\bar \mu _i^{\cal T}({k_i})}}$ is the state of the  $\bar \mu _i^{\cal T}({k_i})$-th leader by the $i$-th follower,  $\bar S_i^R$ and $\bar D_i^R$ are the stack columns of ${S_{\bar \mu _i^{\cal T}({k_i})}}$ and ${D_{\bar \mu _i^{\cal T}({k_i})}}$  for ${k_i} = 1,...,{\bar {\bar l}_i}$, i.e., $\bar S_i^R = col({S_{\bar \mu _i^{\cal T}(1)}},...,{S_{\bar \mu _i^{\cal T}({{\bar {\bar l}}_i})}})$ and $\bar D_i^R = col({D_{\bar \mu _i^{\cal T}(1)}},...,{D_{\bar \mu _i^{\cal T}({{\bar {\bar l}}_i})}})$, where ${S_{\bar \mu _i^{\cal T}({k_i})}}$ and ${D_{\bar \mu _i^{\cal T}({k_i})}}$ are the dynamics of the $\bar \mu _i^{\cal T}({k_i})$- th leader. Moreover, $\eta _j^i = col(\eta _j^{\bar \mu _i^{\cal T}(1)},...,\eta _j^{\bar \mu _i^{\cal T}({{\bar {\bar l}}_i})})$, $S_j^i = col(S_j^{\bar \mu _i^{\cal T}(1)},...,S_j^{\bar \mu _i^{\cal T}({{\bar {\bar l}}_i})})$ and $D_j^i = col(D_j^{\bar \mu _i^{\cal T}(1)},...,D_j^{\bar \mu _i^{\cal T}({{\bar {\bar l}}_i})})$,  $\forall j \in N_i\backslash N_i^L$, where for ${k_i} = 1,...,{\bar {\bar l}_i}$
\begin{align}
\eta _j^{\bar \mu _i^{\cal T}({k_i})} = \left\{ {\begin{array}{*{20}{c}}
{\eta _i^{\bar \mu _i^{\cal T}({k_i})}}\\
{\eta _j^{\bar \mu _i^{\cal T}({k_i})}}
\end{array}} \right.\,\,\,\begin{array}{*{20}{c}}
{\bar \mu _i^{\cal T}({k_i}) \notin \bar N_j^L}\\
{\bar \mu _i^{\cal T}({k_i}) \in \bar N_j^L}
\end{array}, \label{eq:200*} \\
S_j^{\bar \mu _i^{\cal T}({k_i})} = \left\{ {\begin{array}{*{20}{c}}
{S_i^{\bar \mu _i^{\cal T}({k_i})}}\\
{S_j^{\bar \mu _i^{\cal T}({k_i})}}
\end{array}} \right.\,\,\,\begin{array}{*{20}{c}}
{\bar \mu _i^{\cal T}({k_i}) \notin \bar N_j^L}\\
{\bar \mu _i^{\cal T}({k_i}) \in \bar N_j^L}
\end{array}, \label{eq:201*} \\
D_j^{\bar \mu _i^{\cal T}({k_i})} = \left\{ {\begin{array}{*{20}{c}}
{D_i^{\bar \mu _i^{\cal T}({k_i})}}\\
{D_j^{\bar \mu _i^{\cal T}({k_i})}}
\end{array}} \right.\,\,\,\begin{array}{*{20}{c}}
{\bar \mu _i^{\cal T}({k_i}) \notin \bar N_j^L}\\
{\bar \mu _i^{\cal T}({k_i}) \in \bar N_j^L}
\end{array}. \label{eq:202*}
\end{align}

We call ${\eta _i}$, ${\bar S_i}$ and ${\bar D_i}$ as the state and dynamics estimation of the virtual exo-system (\ref{eq:85}) by the $i$-th follower, respectively. 

{\bf Definition 11.}\label{def:12}
Define $\bar {\mathcal{G}}_\lambda ^S = (\bar {\mathcal{V}}_\lambda ^S,\bar {\mathcal{E}} _\lambda ^S,{\mathcal{A}}_\lambda ^S)$ as the sub-graph of reachable agents from the $\lambda $-th leader, $\forall \lambda  \in {\mathcal{R}}$, which is obtained from $\mathcal{G}$ by removing the nodes that are not reachable from the $\lambda $-th leader along with their corresponding edges. That is, $\bar {\mathcal{V}}_\lambda ^S = \bar N_\lambda ^S \cup \left\{ \lambda  \right\}$, where
\begin{align}
\bar N_\lambda ^S = \left\{ {i \in {\mathcal{F}}:\lambda  \in \bar N_i^L} \right\} \label{eq:95}
\end{align}
is the set of followers that are reachable from the $\lambda $-th leader. Moreover,
$\bar \mu _\lambda ^S: = \vec {\bar N}_\lambda ^S$
is the sort function of well-ordered set $(\bar N_\lambda ^S, \le )$, and $l_\lambda ^S = \left| {\bar N_\lambda ^S} \right|$ is the number of followers that are reachable from the $\lambda $-th leader. Furthermore, $\bar {\mathcal{A}}_\lambda ^S$ and $\bar {\mathcal{D}}_\lambda ^S$ are defined as the weighted adjacency matrix and in-degree matrix of the sub-graph $\bar {\mathcal{G}}_\lambda ^S$, respectively, and the Laplacian matrix ${\mathcal{L}}_\lambda ^S$ associated with it, is defined as
${\mathcal{L}}_\lambda ^S = \bar {\mathcal{D}}_\lambda ^S - \bar {\mathcal{A}}_\lambda ^S $.


{\bf Theorem 1.}\label{theorem:3}
Consider the leader dynamics (\ref{eq:7}) and the adaptive distributed observer (\ref{eq:92})-(\ref{eq:94}). Let ${\tilde S_i} = {S_i} - \bar S_i^R$ and ${\tilde D_i} = {D_i} - \bar D_i^R$ be the $i$-th follower's virtual exo-system dynamics estimation errors, and ${\tilde \eta _i} = {\eta _i} - \Omega _i^R$ be the state estimation error of the $i$-th follower's virtual exo-system, $\forall i \in  {\mathcal{F}}$. Then, for any initial conditions ${\tilde S_i}(0)$, ${\tilde D_i}(0)$, and ${\tilde \eta _i}(0)$, one obtains
\begin{enumerate} 
\item For any positive constant $\beta _{}^S$, $\forall i \in  F$, $\mathop {\lim }\limits_{t \to \infty } {\tilde S_i}(t) = 0$, exponentially; 
\item For any positive constant $\beta _{}^D$, $\forall i \in  {\mathcal{F}}$, $\mathop {\lim }\limits_{t \to \infty } {\tilde D_i}(t) = 0$, exponentially; 
\item For any positive constant $\beta _{}^\eta $, and  $\beta _{}^S$, $\forall i \in  {\mathcal{F}}$, $\mathop {\lim }\limits_{t \to \infty } {\tilde \eta _i}(t) = 0$, exponentially.
\end{enumerate}

\textbf{Proof.} 
The proof has three parts. Part (1). Let ${\tilde S^R} = {S^R} - {\bar S^R}$ where ${S^R} = col({S_1},...,{S_n})$, ${\bar S^R} = col(\bar S_1^R,...,\bar S_n^R)$ be the global virtual exo-systems dynamics estimation errors. Using (\ref{eq:93}), one can see that the dynamics of the global virtual exo-systems dynamics estimation errors, i.e., ${\dot {\tilde S}^R}$, is a stack column vector of $\dot {\tilde S}_i^{\bar \mu _i^{\cal T}({k_i})}$, $\forall {k_i} = 1,...,{\bar {\bar l}_i}$, $\forall i \in {\mathcal {F}}$, given by
\begin{align}
\dot {\tilde S}_i^{\bar \mu _i^{\cal T}({k_i})} = \beta _{}^S(\sum\limits_{j \in N_i\backslash N_i^L} {{a_{ij}}} (S_j^{\bar \mu _i^{\cal T}({k_i})} - S_i^{\bar \mu _i^{\cal T}({k_i})}) + \nonumber \\
 {a_{i\bar \mu _i^{\cal T}({k_i})}}({S_{\bar \mu _i^{\cal T}({k_i})}} - S_i^{\bar \mu _i^{\cal T}({k_i})})). \label{eq:98}
\end{align}

Using Definition~11, let the global $\lambda $-th leader dynamics estimation error from the $\lambda $-th leader's reachability perspective, i.e., $\bar {\mathcal {G}}_\lambda ^S$, be
\begin{align}
{\tilde S^\lambda } = {S^\lambda } - ({1_{l_\lambda ^S}} \otimes {S_\lambda }) \label{eq:203*}
\end{align}
where ${S^\lambda } = col(S_{\bar \mu _\lambda ^S(1)}^\lambda ,...,S_{\bar \mu _\lambda ^S(l_\lambda ^S)}^\lambda )$.

Using (\ref{eq:98}) for $\tilde S_i^{\bar \mu _i^{\cal T}({k_i})}$ with $\bar \mu _i^{\cal T}({k_i}) = \lambda $, for $i = \bar \mu _\lambda ^S(1),...,\bar \mu _\lambda ^S(l_\lambda ^S)$, and the fact that $\lambda  \in \bar N_j^L$, $\forall j \in \bar N_\lambda ^S$, the dynamics of the global $\lambda $-th leader dynamics estimation error from the $\lambda $-th leader's reachability perspective, can be written as
\begin{align}
{\dot {\tilde S}^\lambda } =  - {\beta ^S}(H_\lambda ^S \otimes {I_q}){\tilde S^\lambda } \label{eq:99}
\end{align}
where $H_\lambda ^S: = L_\lambda ^S + \delta _\lambda ^S \in {\Re ^{l_\lambda ^S \times l_\lambda ^S}}$, 
${\tilde S^\lambda } = col(\tilde S_{\bar \mu _\lambda ^S(1)}^\lambda ,...,\tilde S_{\bar \mu _\lambda ^S(l_\lambda ^S)}^\lambda ) \in {\Re ^{l_\lambda ^Sq \times q}}$ and $\delta _\lambda ^S = diag({a_{\bar \mu _\lambda ^S(1)\lambda }},\allowbreak ...,{a_{\bar \mu _\lambda ^S(l_\lambda ^S)\lambda }})$.

It follows from the definitions of $\bar N_i^L$ and $\bar N_\lambda ^S$ that there exists at least a directed path from the leader $\lambda $ to $i$-th follower $\forall i \in \bar N_\lambda ^S$ in the sub-graph $\bar {\mathcal {G}}_\lambda ^S$, i.e., Assumption~1 is satisfied for the sub-graph $\bar {\mathcal {G}}_\lambda ^S$. So, based on Lemma 5 of \cite{hu2007leader}, all the eigenvalues of $H_\lambda ^S$ have positive real parts. one has $\dot {\tilde S} =  - {\beta ^S}({H^S} \otimes {I_q})\tilde S$
where $\tilde S = col({\tilde S^{n + 1}},...,{\tilde S^{n + m}}) \in {\Re ^{\sum\nolimits_{\lambda  = n + 1}^{n + m} {l_\lambda ^S} q \times q}}$ is the global virtual exo-system dynamics estimation errors reordered based on leaders' labels and ${H^S} = diag(H_{n + 1}^S,...,H_{n + m}^S) \in {\Re ^{\sum\nolimits_{\lambda  = n + 1}^{n + m} {l_\lambda ^S}  \times \sum\nolimits_{\lambda  = n + 1}^{n + m} {l_\lambda ^S} }}$. 

Appling the vec-operator on the matrix $\dot {\tilde S}$, one has $vec(\dot {\tilde S}) =  - {\beta ^S}({I_q} \otimes {H^S} \otimes {I_q})vec(\tilde S)$.
Since all the eigenvalues of $H_\lambda ^S$, $\forall \lambda  \in \mathcal {R}$ have positive real parts, for any positive constant $\beta _{}^S$, ${\lim _{t \to \infty }}vec(\tilde S(t)) = 0$, exponentially, and consequently ${\lim _{t \to \infty }}vec({\tilde S^\lambda }(t)) = 0$, $\forall \lambda  \in \mathcal {R}$, exponentially. Therefore,
\begin{align}
{\lim _{t \to \infty }}vec(\tilde S_{\bar \mu _\lambda ^S({k_\lambda })}^\lambda (t)) = 0,\forall \lambda  \in \mathcal {R} \label{eq:204*}
\end{align}
and $\forall {k_\lambda } = 1,...,l_\lambda ^S$, exponentially.

Let ${P^S} \in {\Re ^{n \times \sum\nolimits_{\lambda  = n + 1}^{n + m} {l_\lambda ^S} }}$ be a permutation matrix, which permutes the sequence of rows in $\tilde S$ such that ${\tilde S^R} = ({P^S} \otimes {I_q})\tilde S$
where ${\tilde S^R} = col({\tilde S_i},...,{\tilde S_n}) \in {\Re ^{\sum\nolimits_{\lambda  = n + 1}^{n + m} {l_\lambda ^S} q \times q}}$ is the global virtual exo-systems dynamics estimation errors. Taking the derivative and using vector operator yields $vec({\dot {\tilde S}^R}) =  - {\beta ^S}({I_q} \otimes {P^S}{H^S}{({P^S})^T} \otimes {I_q})vec({\tilde S^R})$.
Note that, a permutation matrix is always nonsingular, and ${P^S}{({P^S})^T} = {I_n}$. Therefore, recalling that under Assumption~1 all the eigenvalues of $H_\lambda ^S$, $\forall \lambda  \in \mathcal {R}$  have positive real parts, $ - ({P^S} \otimes {I_q})({H^S} \otimes {I_q})({({P^S})^T} \otimes {I_q})$ is Hurwitz. This implies that ${\lim _{t \to \infty }}vec(\tilde S_i^{\bar \mu _i^{\cal T}({k_i})}(t)) = 0$, $\forall i \in \mathcal {F}$, and $\forall {k_i} = 1,...,{\bar {\bar l}_i}$, exponentially, and ${\lim _{t \to \infty }}vec({\tilde S_i}(t)) = 0$, $\forall i \in \mathcal {F}$, which yields $\mathop {\lim }\limits_{t \to \infty } {\tilde S_i}(t) = 0$, $\forall i \in \mathcal {F}$, exponentially. 

Part (2). Similar to part (1), one can observe that $\mathop {\lim }\limits_{t \to \infty } {\tilde D_i}(t) = 0$ exponentially, $\forall i \in \mathcal {F}$ for any positive constant $\beta _{}^D$.

Part (3). Let $\tilde \eta  = \eta  - \Omega $ where $\eta  = col({\eta _1},...,{\eta _n})$, $\Omega  = col(\Omega _1^R,...,\Omega _n^R)$ be the global virtual exo-systems state estimation errors. Using (\ref{eq:92}), one can see that the dynamics of the global virtual exo-systems state estimation errors, i.e., $\dot {\tilde \eta} $, is a stack column vector of $\dot {\tilde \eta} _i^{\bar \mu _i^{\cal T}({k_i})}$, $\forall {k_i} = 1,...,{\bar {\bar l}_i}$, $\forall i \in \mathcal {F}$, as
\begin{align}
\begin{array}{l}
\dot {\tilde \eta} _i^{\bar \mu _i^{\cal T}({k_i})} = {S_{\bar \mu _i^{\cal T}({k_i})}}\tilde \eta _i^{\bar \mu _i^{\cal T}({k_i})} + \tilde S_i^{\bar \mu _i^{\cal T}({k_i})}\tilde \eta _i^{\bar \mu _i^{\cal T}({k_i})} + \tilde S_i^{\bar \mu _i^{\cal T}({k_i})}{\omega _{\bar \mu _i^{\cal T}({k_i})}}\\
\,\,\,\,\,\, + \beta _{}^\eta (\sum\limits_{j \in N_i\backslash N_i^L} {{a_{ij}}} (\tilde \eta _j^{\bar \mu _i^{\cal T}({k_i})} - \tilde \eta _i^{\bar \mu _i^{\cal T}({k_i})}) - {a_{i\bar \mu _i^{\cal T}({k_i})}}\tilde \eta _i^{\bar \mu _i^{\cal T}({k_i})})
\end{array} \label{eq:105}
\end{align}
where $\tilde \eta _i^{\bar \mu _i^{\cal T}({k_i})} = \eta _i^{\bar \mu _i^{\cal T}({k_i})} - {\omega _{\bar \mu _i^{\cal T}({k_i})}},\forall {k_i} = 1,...,{\bar {\bar l}_i}$
and
\begin{align}
\tilde \eta _j^{\bar \mu _i^{\cal T}({k_i})} = \left\{ {\begin{array}{*{20}{c}}
{\tilde \eta _i^{\bar \mu _i^{\cal T}({k_i})}}\\
{\eta _j^{\bar \mu _i^{\cal T}({k_i})} - {\omega _{\bar \mu _i^{\cal T}({k_i})}}}
\end{array}\,\,\,} \right.\,\,\,\begin{array}{*{20}{c}}
{\bar \mu _i^{\cal T}({k_i}) \notin \bar N_j^L}\\
{\bar \mu _i^{\cal T}({k_i}) \in \bar N_j^L}
\end{array} \label{eq:206*}
\end{align}

Using Definition~11, let the global $\lambda $-th leader state estimation error from the $\lambda $-th leader's reachability perspective, i.e., $\bar {\mathcal {G}}_\lambda ^S$, be ${\tilde \eta ^\lambda } = {\eta ^\lambda } - ({1_{l_\lambda ^S}} \otimes {\omega _\lambda })$
where ${\eta ^\lambda } = col(\eta _{\bar \mu _\lambda ^S(1)}^\lambda ,...,\eta _{\bar \mu _\lambda ^S(l_\lambda ^S)}^\lambda )$. Based on (\ref{eq:94}),  ${\dot {\tilde \eta} ^\lambda }$  can be written as
\begin{align}
{\dot {\tilde \eta} ^\lambda } = ({I_{l_\lambda ^S}} \otimes {S_\lambda } - \beta _{}^\eta (H_\lambda ^S \otimes {I_q})){\tilde \eta ^\lambda } + {\bar {\tilde S}^\lambda }{\tilde \eta ^\lambda } + {\bar {\tilde S}^\lambda }({1_{l_\lambda ^S}} \otimes {\omega _\lambda }) \label{eq:106}
\end{align}
where ${\tilde \eta ^\lambda } = col(\tilde \eta _{\bar \mu _\lambda ^S(1)}^\lambda ,...,\tilde \eta _{\bar \mu _\lambda ^S(l_\lambda ^S)}^\lambda ) \in {\Re ^{l_\lambda ^Sq \times 1}}$, ${\bar {\tilde S}^\lambda } = diag({\tilde S^\lambda })$, and $H_\lambda ^S$ is given in (\ref{eq:203*}). Note that $\lambda  \in \bar N_j^L$, $\forall j \in \bar N_\lambda ^S$, which implies $\tilde \eta _j^\lambda  = \eta _j^\lambda  - {\omega _\lambda }$ based on (\ref{eq:206*}).

According to Assumption 4, and recalling that all the eigenvalues of $H_\lambda ^S$ have positive real parts, one can see that for any positive constant $\beta _{}^\eta $, and $\beta _{}^S$, the matrix $({I_{l_\lambda ^S}} \otimes {S_\lambda } - \beta _{}^\eta (H_\lambda ^S \otimes {I_q}))$ is Hurwitz. Moreover, based on part (1), ${\bar {\tilde S}^\lambda }({1_{l_\lambda ^S}} \otimes {\omega _\lambda }) \to {0_{l_\lambda ^Sq \times 1}}$, exponentially.

The dynamics of the global virtual exo-system state estimation errors reordered based on the leaders' labels, denoted by ${\tilde \eta ^R}$, can be written as ${\dot {\tilde \eta} ^R} = \Theta {\tilde \eta ^R} + \bar {\tilde S}{\tilde \eta ^R} + \bar {\tilde S}{\omega ^R}$
where ${\tilde \eta ^R} = col({\tilde \eta ^{n + 1}},...,{\tilde \eta ^{n + m}}) \in {\Re ^{\sum\nolimits_{\lambda  = n + 1}^{n + m} {l_\lambda ^S} q \times 1}}$, ${\omega ^R} = col(({1_{l_{n + 1}^S}} \otimes {\omega _{n + 1}}),...,({1_{l_{n + m}^S}} \otimes {\omega _{n + m}})) \in {\Re ^{\sum\nolimits_{\lambda  = n + 1}^{n + m} {l_\lambda ^S} q \times 1}}$,  $\Theta  = diag({\Theta _{n + 1}},...,{\Theta _{n + m}})$, ${\Theta _k} = ({I_{l_k^S}} \otimes {S_k} - \beta _{}^\eta (H_k^S \otimes {I_q}))$, $\forall k \in \mathcal {R}$, and $\bar {\tilde S} = diag({\bar {\tilde S}^{n + 1}},...,{\bar {\tilde S}^{n + m}})$. Moreover, based on the part (1),   , exponentially. Moreover, based on the part (1), $\bar {\tilde S}{\omega ^R} \to {0_{^{\sum\nolimits_{\lambda  = n + 1}^{n + m} {l_\lambda ^S} q \times 1}}}$, exponentially.

So, based on Lemma 5 of \cite{cai2017adaptive}, for any positive constant $\beta _{}^\eta$ and $\beta _{}^S$, $\forall i \in \mathcal {F}$, ${\lim _{t \to \infty }}{\tilde \eta ^R}(t) = 0$ exponentially, and consequently ${\lim _{t \to \infty }}{\tilde \eta ^\lambda }(t) = 0$, $\forall \lambda  \in \mathcal {R}$, exponentially. Therefore, ${\lim _{t \to \infty }}\tilde \eta _{\bar \mu _\lambda ^S({k_\lambda })}^\lambda (t) = 0$, $\forall \lambda  \in \mathcal {R}$, and $\forall {k_\lambda } = 1,...,l_\lambda ^S$, exponentially.  

Let ${P^\eta } \in {\Re ^{\sum\nolimits_{\lambda  = n + 1}^{n + m} {l_\lambda ^S}  \times \sum\nolimits_{\lambda  = n + 1}^{n + m} {l_\lambda ^S} }}$ be a permutation matrix, which permutes the sequence of rows in ${\tilde \eta ^R}$ such that $\tilde \eta  = ({P^\eta } \otimes {I_q}){\tilde \eta ^R}$
where $\tilde \eta  = col({\tilde \eta _1},...,{\tilde \eta _n}) \in {\Re ^{\sum\nolimits_{\lambda  = n + 1}^{n + m} {l_\lambda ^S} q \times 1}}$ is the global virtual exo-systems state estimation errors.

The state of the global virtual exo-system dynamics estimation errors, i.e., $\dot {\tilde \eta} $, can be written as
\begin{align}
\dot {\tilde \eta}  = ({P^\eta } \otimes {I_q})\Theta ({({P^\eta })^T} \otimes {I_q})\tilde \eta  + ({P^\eta } \otimes {I_q})\bar {\tilde S}\tilde \eta 
 + ({P^\eta } \otimes {I_q})\bar {\tilde S}{\omega ^R} \label{eq:109}
\end{align}
where ${\omega ^R} = col(({1_{l_{n + 1}^S}} \otimes {\omega _{n + 1}}),...,({1_{l_{n + m}^S}} \otimes {\omega _{n + m}})) \in {\Re ^{\sum\nolimits_{\lambda  = n + 1}^{n + m} {l_\lambda ^S} q \times 1}}$, $\Theta  = diag({\Theta _{n + 1}},...,{\Theta _{n + m}})$, ${\Theta _k} = ({I_{l_k^S}} \otimes {S_k} - \beta _{}^\eta (H_k^S \otimes {I_q}))$, $\forall k \in \mathcal {R}$, and $\bar {\tilde S} = diag({\bar {\tilde S}^{n + 1}},...,{\bar {\tilde S}^{n + m}})$. 

Note that, a permutation matrix is always nonsingular, and ${P^\eta }{({P^\eta })^T} = {I_{\sum\nolimits_{\lambda  = n + 1}^{n + m} {l_\lambda ^S} }}$. According to Assumption 4, and recalling that all the eigenvalues of $H_\lambda ^S$ have positive real parts, one can see that for any positive constant $\beta _{}^\eta $, and $\beta _{}^S$, the matrix $({I_{l_\lambda ^S}} \otimes {S_\lambda } - \beta _{}^\eta (H_\lambda ^S \otimes {I_q}))$, $\lambda  \in \mathcal {R}$ is Hurwitz and consequently, the matrix $({P^\eta } \otimes {I_q})\Theta ({({P^\eta })^T} \otimes {I_q})$  is Hurwitz. Moreover, based on the part (1), $({P^\eta } \otimes {I_q})\bar {\tilde S}{\Omega _\eta } \to {0_{^{nq \times 1}}}$, exponentially. So, based on Lemma 5 of \cite{cai2017adaptive}, for any positive constant $\beta _{}^\eta $ and $\beta _{}^S$, $\forall i \in \mathcal {F}$, ${\lim _{t \to \infty }}\tilde \eta (t) = 0$ exponentially, and consequently ${\lim _{t \to \infty }}\tilde \eta _i^{\bar \mu _i^{\cal T}({k_i})}(t) = 0$ and ${\lim _{t \to \infty }}{\tilde \eta _i}(t) = 0$, $\forall i \in \mathcal {F}$, and $\forall {k_i} = 1,...,{\bar {\bar l}_i}$ exponentially. This completes the proof. \hfill {$\square$}


{\bf Corollary 1.}\label{cor:2}
Consider the leader dynamics (\ref{eq:7}) and the adaptive distributed observer (\ref{eq:92}) -(\ref{eq:94}). Let $\tilde y_i^*(t) = (\Phi _{P\ell }^i \otimes {I_Q}){\bar D_i}{\eta _i} - y_i^*$ be the output estimation error of the $i$-th virtual exo-system dynamics, where $y_i^*$, is given in (\ref{eq:85}) and ${\bar D_i} = diag(D_i^{\bar \mu _i^{\cal T}(1)}, \ldots ,D_i^{\bar \mu _i^{\cal T}({{\bar {\bar l}}_i})})$, $\forall i \in {\mathcal{F}}$. Then, for any positive constant $\beta _{}^\eta $, $\beta _{}^S$, and $\beta _{}^D$, and any initial condition, $\tilde y_i^*(0)$, $\forall i \in {{\mathcal{F}}}$, $\mathop {\lim }\limits_{t \to \infty } \tilde y_i^*(t) = 0$, exponentially fast. 

 \textbf{Proof.} The proof is an immediate consequence of Theorem 1 and is thus omitted. \hfill {$\square$}
\vspace{0.1in}

 {\textbf{Remark 3.}  It worth noting that the distributed observer approach is a promising approach to deal with various leader-following and containment control problems of multiagent systems such as cooperative output regulation problems \cite{su2011cooperative,su2012cooperative,haghshenas2015containment,seyboth2016cooperative} and the leader-following output synchronization problem \cite{zhu2015general}. The authors in \cite{cai2017adaptive} introduced an adaptive distributed observer approach which obviates the requirement of knowing the dynamics of the identical leaders which makes it more realistic for some applications. Inspired by this standard notion, the proposed adaptive distributed observer in this paper estimates the dynamics and states of the influential leaders in a fully distributed fashion without significantly increase the communication burden.}

\section{Distributed solution to containment control problem with heterogeneous leaders} \label{Sec:6}

In this section, a distributed dynamic output feedback control protocol is introduced to solve the output containment control problem with heterogeneous leaders.


{\bf Assumption 5.}\label{as:7}
The linear matrix equations
\begin{align}
\begin{array}{l}
\Pi _i^R\bar {\bar S}_i^R = {A_i}\Pi _i^R + {B_i}\Gamma _i^R\\
0 = {C_i}\Pi _i^R - (\Phi _{P\ell }^i \otimes {I_Q})\bar {\bar D}_i^R
\end{array} \label{eq:117}
\end{align}
have solutions, where $\bar {\bar S}_i^R = diag({S_{\bar \mu _i^{\cal T}(1)}},...,{S_{\bar \mu _i^{\cal T}({{\bar {\bar l}}_i})}})$, $\bar {\bar D}_i^R = diag({D_{\bar \mu _i^{\cal T}(1)}},...,{D_{\bar \mu _i^{\cal T}({{\bar {\bar l}}_i})}})$, $\Pi _i^R \in {\Re ^{{N_i} \times {{\bar {\bar l}}_i}q}}$, and $\Gamma _i^R \in {\Re ^{{P_i} \times {{\bar {\bar l}}_i}q}}$, $\forall i \in {\mathcal{F}}$.

By solving Problem~2, the output containment control problem~1 is also solved. To solve Problem~2, the following state feedback control is introduced in this paper.
\begin{align}
{u_i} = K_i^1{x_i} + K_i^2{\eta _i} \label{eq:118}
\end{align}
where ${\eta _i}$ is given by (\ref{eq:92}), and $K_i^1 \in {\Re ^{{P_i} \times {N_i}}}$ and $K_i^2 \in {\Re ^{{P_i} \times {{\bar {\bar l}}_i}q}}$ are design feedback and feedforward gain matrices, respectively, for agent  $i$, $i \in {\mathcal{F}}$.

Choosing sufficiently large $\beta _{}^S > 0$ and $\beta _{}^D > 0$ makes the observers (\ref{eq:93})-(\ref{eq:94}) converges sufficiently fast, and thus we can assume ${S_i} \to \bar S_i^R$ and ${D_i} \to \bar D_i^R$, which  implies that ${\bar S_i} \to \bar {\bar S}_i^R$ and ${\bar D_i} \to \bar {\bar D}_i^R$. One can now write ${\dot \eta _i}$ in (\ref{eq:92}) as the stack column vector $\dot \eta _i^{\bar \mu _i^{\cal T}({k_i})}$, $\forall {k_i} = 1,...,{\bar {\bar l}_i}$, where
\begin{align}
\dot \eta _i^{\bar \mu _i^{\cal T}({k_i})} =& {S_{\bar \mu _i^{\cal T}({k_i})}}\eta _i^{\bar \mu _i^{\cal T}({k_i})} + \beta _{}^\eta  (\sum\limits_{j \in N_i\backslash N_i^L} {{a_{ij}}} (\eta _j^{\bar \mu _i^{\cal T}({k_i})}- \nonumber \\
& \eta _i^{\bar \mu _i^{\cal T}({k_i})}) + {a_{i\bar \mu _i^{\cal T}({k_i})}}({\omega _{\bar \mu _i^{\cal T}({k_i})}} - \eta _i^{\bar \mu _i^{\cal T}({k_i})})). \label{eq:111}
\end{align}

Based on (\ref{eq:111}) and Definition~11, the dynamics of the global $\lambda $-th leader state estimation from the $\lambda $-th leader's reachability perspective, denote by ${\dot \eta ^\lambda }$, can be written as
\begin{align}
{\dot \eta ^\lambda } = ({I_{l_\lambda ^S}} \otimes {S_\lambda } - \beta _{}^\eta (H_\lambda ^S \otimes {I_q})){\eta ^\lambda } + \beta _{}^\eta (H_\lambda ^S \otimes {I_q})({1_{l_\lambda ^S}} \otimes {\omega _\lambda }) \label{eq:112}
\end{align}
where ${\eta ^\lambda } = col(\eta _{\bar \mu _\lambda ^S(1)}^\lambda ,...,\eta _{\bar \mu _\lambda ^S(l_\lambda ^S)}^\lambda ) \in {\Re ^{l_\lambda ^Sq \times 1}}$, $\forall \lambda  \in {\mathcal{R}}$, with $\bar \mu _\lambda ^S = \vec {\bar N}_\lambda ^S$, $\bar N_\lambda ^S = \left\{ {i \in {\mathcal{F}}:\lambda  \in \bar N_i^L} \right\}$, and $l_\lambda ^S = \left| {\bar N_\lambda ^S} \right|$ given in (\ref{eq:95}). One has
\begin{align}
{\dot \eta ^R} = \Theta {\eta ^R} + \bar \Theta {\omega ^R} \label{eq:113}
 \end{align}
 where ${\eta ^R} = col({\eta ^{n + 1}},...,{\eta ^{n + m}}) \in {\Re ^{\sum\nolimits_{\lambda  = n + 1}^{n + m} {l_\lambda ^S} q \times 1}}$, ${\omega ^R} = col(({1_{l_{n + 1}^S}} \otimes {\omega _{n + 1}}),...,({1_{l_{n + m}^S}} \otimes {\omega _{n + m}})) \in {\Re ^{\sum\nolimits_{\lambda  = n + 1}^{n + m} {l_\lambda ^S} q \times 1}}$, $\Theta  = diag({\Theta _{n + 1}},...,{\Theta _{n + m}})$, $\bar \Theta  = diag({\bar \Theta _{n + 1}},...,{\bar \Theta _{n + m}})$, ${\Theta _k} = ({I_{l_k^S}} \otimes {S_k} - \beta _{}^\eta (H_k^S \otimes {I_q}))$ and ${\bar \Theta _k} = \beta _{}^\eta (H_k^S \otimes {I_q})$, $\forall k \in {\mathcal{R}}$.

After some manipulation and using a linear transformation, (\ref{eq:113}) can be rewritten as
\vspace{-0.1in}
\begin{align}
\dot \eta  = ({P^\eta } \otimes {I_q})\Theta ({({P^\eta })^T} \otimes {I_q})\eta  + \beta _{}^\eta \Omega \label{eq:114}
\end{align}
 with $\eta  = ({P^\eta } \otimes {I_q}){\eta ^R}$ and $\Omega  = ({P^\eta } \otimes {I_q}){\omega ^R}$ 
where $\eta  = col({\eta _1},...,{\eta _n})$, ${\eta _i} = col(\eta _i^{\bar \mu _i^{\cal T}(1)},...,\eta _i^{\bar \mu _i^{\cal T}({{\bar {\bar l}}_i})})$, $i \in {\mathcal{F}}$, $\Omega  = col(\Omega _1^R,...,\Omega _n^R)$, and ${P^\eta } \in {\Re ^{\sum\nolimits_{\lambda  = n + 1}^{n + m} {l_\lambda ^S}  \times \sum\nolimits_{\lambda  = n + 1}^{n + m} {l_\lambda ^S} }}$ is a permutation matrix, which permutes the sequence of rows in ${\eta ^R}$ and ${\omega ^R}$. 
 
 Utilizing (\ref{eq:114}), the composition of (\ref{eq:7})-(\ref{eq:8}), the control law (\ref{eq:118}) along with distributed observers (\ref{eq:92})-(\ref{eq:94}), the virtual exo-system (\ref{eq:85}), and local containment error (\ref{eq:86}) results in the following closed-loop systems
 \begin{align}
\dot \Omega  &= \bar {\bar S}_{}^R\Omega \label{eq:119} \\
{\dot X_C} &= {A_C}{X_C} + {B_C}\Omega  \label{eq:120} \\
E &= {C_C}{X_C} + {D_C}\Omega  \label{eq:121}
 \end{align}
with
\vspace{-0.1in}
\begin{align}
{A_C} &= \left[ {\begin{array}{*{20}{c}}
{A + BK_{}^1}&{BK_{}^2}\\
{{0_{\sum\nolimits_{\lambda  = n + 1}^{n + m} {l_\lambda ^S} q \times \sum\nolimits_{i = 1}^n {{N_i}} }}}&{(P \otimes {I_q})\Theta ({P^T} \otimes {I_q})}
\end{array}} \right], \nonumber \\
{B_C} &= \left[ {\begin{array}{*{20}{c}}
{{0_{\sum\nolimits_{i = 1}^n {{N_i}}  \times \sum\nolimits_{i = 1}^n {{{\bar {\bar l}}_i}q} }}}\\
{\beta _{}^\eta {I_{\sum\nolimits_{i = 1}^n {{{\bar {\bar l}}_i}q} }}}
\end{array}} \right], 
{C_C} = \left[ {\begin{array}{*{20}{c}}
C&{{0_{nQ \times \sum\nolimits_{\lambda  = n + 1}^{n + m} {l_\lambda ^S} q}}}
\end{array}} \right], \nonumber \\
{D_C}& =  - ({\Phi ^R} \otimes {I_Q})\bar {\bar D}_{}^R,
 \label{eq:122}
\end{align}
 where ${X_C} = col(X,\eta )$, $X = col({x_1},...,{x_n})$, $\eta  = col({\eta _1},...,{\eta _n}) \allowbreak \in {\Re ^{\sum\nolimits_{\lambda  = n + 1}^{n + m} {l_\lambda ^S} q}}$, $\Omega  = col(\Omega _1^R,...,\Omega _n^R) \in {\Re ^{\sum\nolimits_{i = 1}^n {{{\bar {\bar l}}_i}q} }}$, $E = col({\bar e_1},...,{\bar e_n})$, $\bar {\bar S}_{}^R = diag(\bar {\bar S}_1^R,...,\bar {\bar S}_n^R)$, $\bar {\bar D}_{}^R = diag(\bar {\bar D}_1^R,...\allowbreak ,\bar {\bar D}_n^R)$,  $A = diag({A_1},...,{A_n})$, $B = diag({B_1},...,{B_n})$,  $K_{}^1 = diag(K_1^1,...,K_n^1)$, $K_{}^2 = diag(K_1^2,...,K_n^2)$, $C = diag({C_1},...,{C_n})$, and ${\Phi ^R} = diag(\Phi _{P\ell }^1,...,\Phi _{P\ell }^n) \in {\Re ^{n \times \sum\nolimits_{i = 1}^n {{{\bar {\bar l}}_i}} }}$.

 

In Problem~2, $li{m_{t \to \infty }}{\bar e_i}(t) = 0$, $\forall i \in {\mathcal{F}}$ or equivalently ${\lim _{t \to \infty }}E\left( t \right) = 0$, where $E = col({\bar e_1},...,{\bar e_n})$. This implies that the output of each follower should be synchronized to its reference trajectory $y_i^*(t)$, $\forall i \in {\mathcal{F}}$. This is analogous to the output regulation problem \cite{huang2016certainty}. 
Therefore, the following problem can be formulated and be replaced with Problem~2 to solve FHCCP.
 
{\bf Problem 3.}\label{prb:3}
Let Assumptions~1-5 be hold. Consider the multi-agent system (\ref{eq:7})-(\ref{eq:8}) with the digraph $\mathcal{G}$. Design the control protocols given by (\ref{eq:118}) such that the closed-loop system (\ref{eq:119})-(\ref{eq:121}) satisfies the following properties:
\begin{enumerate} 
\item The matrix ${A_C}$ in (\ref{eq:122}) is Hurwitz.
\item For any initial conditions ${x_i}(0)$, ${x_i}(0)$, $\forall i \in {\mathcal{F}}$ and ${\omega _k}(0)$, $\forall k \in {\mathcal{R}}$, ${\lim _{t \to \infty }}E\left( t \right) = 0$, where $E$ is defined in (\ref{eq:121}).
\end{enumerate}

In order to solve Problem~3, we need the following lemma.

{\bf Lemma 5.}\label{lemma:11}
Suppose that Property 1 is fulfilled by the distributed control laws (\ref{eq:118}). Then, ${\lim _{t \to \infty }}E\left( t \right) = 0$, if there exists a matrix ${\bar X}_C$ that satisfies the following linear matrix equations, $\forall k \in {\mathcal{R}}$
\begin{align}
\left\{ \begin{array}{l}
{A_C}{{\bar X}_C} + {B_C} = {{\bar X}_C}\bar {\bar S}_{}^R\\
{C_C}{{\bar X}_C} + {D_C} = {0_{nQ \times \sum\nolimits_{i = 1}^n {{{\bar {\bar l}}_i}q} }}
\end{array} \right. \label{eq:123}
\end{align}
\textbf{Proof.} 
Let ${\tilde X_C} = {X_C} - {\bar X_C}\Omega $. It follows from (\ref{eq:120}), (\ref{eq:121}), (\ref{eq:123}), and some manipulation that ${\dot {\tilde X}_C} = {A_C}{\tilde X_C}$ and $E = {C_C}{\tilde X_C}$.
Since we assumed Property 1 is fulfilled, i.e., ${A_C}$ is Hurwitz, we have ${\lim _{t \to \infty }}{\tilde X_C} = 0$, which implies that ${\lim _{t \to \infty }}E\left( t \right) = 0$. This completes the proof. \hfill {$\square$}
 
It is shown in the following theorem that Problem~3 and consequently Problem~2 and Problem~1 can be solved using the distributed control laws (\ref{eq:118}) along with distributed observer (\ref{eq:92}) -(\ref{eq:94}), and Algorithms 1-3.
 
{\bf Theorem 2.}\label{theorem:4}
 Consider the multi-agent system (\ref{eq:7}) -(\ref{eq:8}). Let Assumptions~1-5 be satisfied. Let $K_i^1$ be chosen such that ${A_i} + {B_i}K_i^1$ is Hurwitz, and $K_i^2$ be given by 
 \vspace{-0.1in}
 \begin{align}
K_i^2 = \Gamma _i^R - K_i^1\Pi _i^R \label{eq:126}
 \end{align}
 where $\Pi _i^R$ and $\Gamma _i^R$, $ i \in {\mathcal{F}}$, are the solutions of (\ref{eq:117}). Then, Problem~3 is solved using the distributed control laws (\ref{eq:118}) along with distributed observers (\ref{eq:92})-(\ref{eq:94}), and Algorithms 1-3, for any positive constants $\beta _{}^\eta$, $\beta _{}^S$, and $\beta _{}^D$.

\textbf{Proof.} 
Under Assumption~1, based on (\ref{eq:109}), for any positive constant $\beta _{}^\eta$ and $\beta _{}^S$, $({P^\eta } \otimes {I_q})\Theta ({({P^\eta })^T} \otimes {I_q})$ is Hurwitz. Moreover, under Assumption~3, there exists a $K_i^1$ such that  ${A_i} + {B_i}K_i^1$ is Hurwitz. Therefore, due to the block-triangular structure of ${A_C}$, ${A_C}$ is Hurwitz, and the observer states, i.e., ${\eta _i}$, is independent of the follower states, i.e., ${x_i}$, $\forall i \in \mathcal {F}$, so based on the separation principle, they can be designed independent of each other. As conclusion, the multi-agent system (\ref{eq:7})-(\ref{eq:8}), under the distributed control laws (\ref{eq:118}) along with distributed observers (\ref{eq:92})-(\ref{eq:94}), satisfies Property 1 in Problem~3. 
 
To complete the proof, it remains to verify Property 2 in Problem~3. Let $K_i^2$ be given by (\ref{eq:126}). Then, under Assumption~5, we obtain
\vspace{-0.1in}
\begin{align}
\Pi \bar {\bar S}_{}^R = (A + BK_{}^1)\Pi  + BK_{}^2 \label{eq:127}
\end{align}
where $\Pi  = diag(\Pi _i^R)$. Set
\begin{align}
{\bar X_C} := \left[ {\begin{array}{*{20}{c}}
{{{(C)}^T}{{(C{{(C)}^T})}^{ - 1}}({\Phi ^R} \otimes {I_Q})\bar {\bar D}_{}^R}\\
{{I_{\sum\nolimits_{i = 1}^n {{{\bar {\bar l}}_i}} }} \otimes {I_q}}
\end{array}} \right]. \label{eq:128}
\end{align}
Using (\ref{eq:122}), (\ref{eq:127}) and (\ref{eq:128}), one has
{small
\begin{align}
&{A_C}{\bar X_C} + {B_C} = \left[ {\begin{array}{*{20}{c}} {BK_{}^2({ I_{\sum\nolimits_{i = 1}^n {{{\bar {\bar l}}_i}} }} \otimes {I_q})} \\ {\beta _{}^\eta {I_{\sum\nolimits_{i = 1}^n {{{\bar {\bar l}}_i}q} }} }\end{array}}\right] + \nonumber \\
& \quad \quad \, \left[ {\begin{array}{*{20}{c}}
{(A + BK_{}^1){{(C)}^T}{{(C{{(C)}^T})}^{ - 1}}({\Phi ^R} \otimes {I_Q})\bar {\bar D}_{}^R }\\
{({P^\eta } \otimes {I_q})\Theta ({{({P^\eta })}^T} \otimes {I_q})({I_{\sum\nolimits_{i = 1}^n {{{\bar {\bar l}}_i}} }} \otimes {I_q}) }
\end{array}} \right]  \nonumber \\  
&\quad = \left[ {\begin{array}{*{20}{c}}
{{{(C)}^T}{{(C{{(C)}^T})}^{ - 1}}({\Phi ^R} \otimes {I_Q})\bar {\bar D}_{}^R\bar {\bar S}_{}^R}\\
{({I_{\sum\nolimits_{i = 1}^n {{{\bar {\bar l}}_i}} }} \otimes {I_q})\bar {\bar S}_{}^R}
\end{array}} \right] = {\bar X_C}\bar {\bar S}_{}^R. \label{eq:129}
\end{align} }
 Note that $({P^\eta } \otimes {I_q})\Theta ({({P^\eta })^T} \otimes {I_q}) = \bar {\bar S}_{}^R - \beta _{}^\eta ({P^\eta } diag(H_{n + 1}^S,...,H_{n + m}^S){({P^\eta })^T} \otimes {I_q}).$
  Furthermore, using (\ref{eq:92}), (\ref{eq:93}) and (\ref{eq:99}), one has ${C_C}{\bar X_C} + {D_C}= C{C^T}{(C{C^T})^{ - 1}}({\Phi ^R} \otimes {I_Q})\bar {\bar D}_{}^R - ({\Phi ^R} \otimes {I_Q})\bar {\bar D}_{}^R = {0_{nQ \times \sum\nolimits_{i = 1}^n {{{\bar {\bar l}}_i}q} }}.$
 Therefore, ${\bar X_C}$ satisfies the equations (\ref{eq:123}) and it follows from Lemma~5 that property 2 is also satisfied, i.e., ${\lim _{t \to \infty }}E\left( t \right) = 0$, which completes the proof.
 \hfill {$\square$}

\section{Simulation} \label{Sec:7}

\begin{figure}
\begin{center}
\includegraphics[height=2.5 cm]{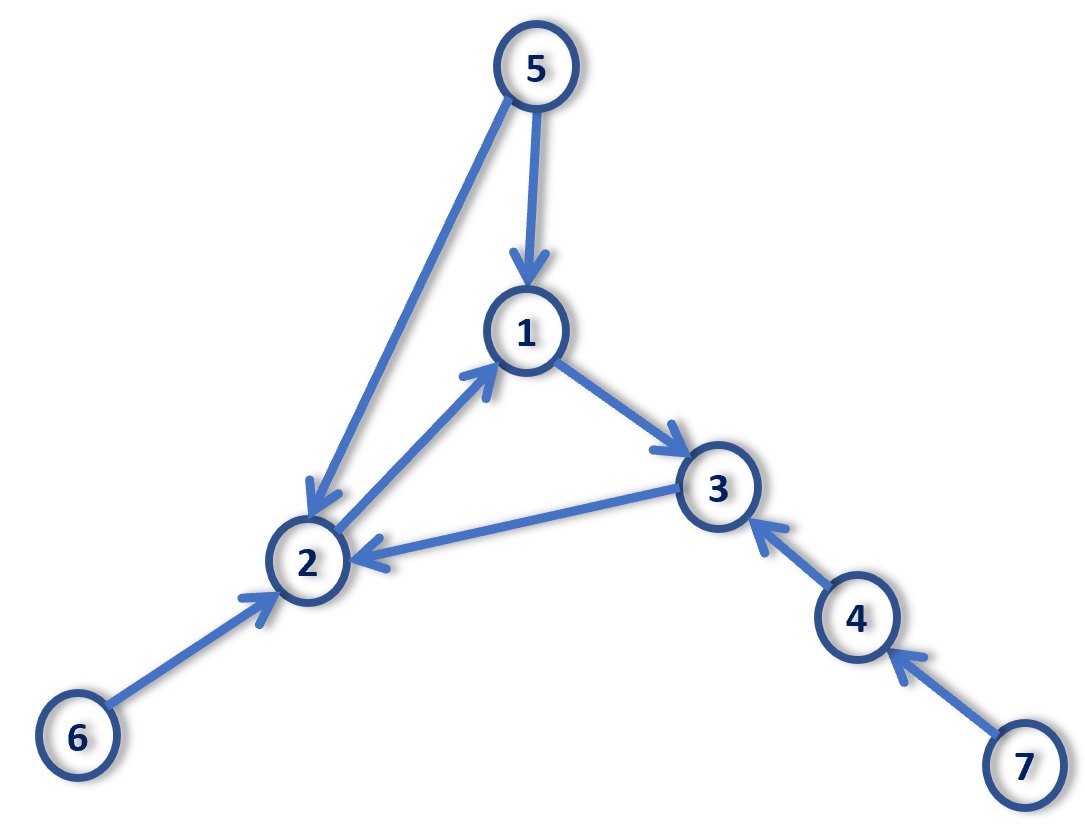}     
\caption{Communication graph.}
\label{fig1}                     
\end{center}                                
\end{figure}

Consider the multi-agent system consist of four heterogeneous followers and three heterogeneous leaders, with the fixed communication graph illustrated in Fig.~\ref{fig1}. All the communication weights are chosen to be one, and nodes $1,2,3$, and $4$ represent the followers and nodes $5, 6$, and $7$ represent three heterogeneous leaders. 
The dynamics of leaders and followers are given as
{ \small
\begin{align}
\begin{array}{l}
\begin{array}{*{20}{l}}
{{S_5} = \left[ {\begin{array}{*{20}{c}}
1&{ - 3}\\
1&{ - 1}
\end{array}} \right],{D_5} = \left[ {\begin{array}{*{20}{c}}
1&0\\
0&1
\end{array}} \right],{S_6} = \left[ {\begin{array}{*{20}{c}}
1&{ - 4}\\
1&{ - 1}
\end{array}} \right],}\\
{{D_6} = \left[ {\begin{array}{*{20}{c}}
1&0\\
0&1
\end{array}} \right],{S_7} = \left[ {\begin{array}{*{20}{c}}
1&{ - 5}\\
1&{ - 1}
\end{array}} \right],{D_7} = \left[ {\begin{array}{*{20}{c}}
1&0\\
0&1
\end{array}} \right],}
\end{array}\\
{A_1} = \left[ {\begin{array}{*{20}{c}}
1&{ - 1}\\
1&0
\end{array}} \right],{B_1} = \left[ {\begin{array}{*{20}{c}}
{ - 2}&{ - 1}\\
1&2
\end{array}} \right],{({C_1})^T} = \left[ {\begin{array}{*{20}{c}}
1&0\\
0&1
\end{array}} \right],
\end{array} \nonumber
\end{align}
\begin{align}
\begin{array}{l}
{A_2} = \left[ {\begin{array}{*{20}{c}}
2&0\\
2&2
\end{array}} \right],{B_2} = \left[ {\begin{array}{*{20}{c}}
{ - 1}&{ - 2}\\
{ - 2}&{ - 1}
\end{array}} \right],{({C_2})^T} = \left[ {\begin{array}{*{20}{c}}
1&0\\
0&1
\end{array}} \right],\\
{A_3} = \left[ {\begin{array}{*{20}{c}}
{ - 1}&0&0\\
0&3&0\\
0&3&2
\end{array}} \right],{B_3} = \left[ {\begin{array}{*{20}{c}}
4&1&1\\
1&4&1\\
1&1&4
\end{array}} \right],{({C_3})^T} = \left[ {\begin{array}{*{20}{l}}
{0\,\,0}\\
{1\,\,0}\\
{0\,\,1}
\end{array}} \right],\\
{A_4} = \left[ {\begin{array}{*{20}{c}}
{ - 1}&0&0\\
0&2&{ - 1}\\
0&3&7
\end{array}} \right],{B_4} = \left[ {\begin{array}{*{20}{c}}
1&0&0\\
0&1&0\\
0&0&1
\end{array}} \right],{({C_4})^T} = \left[ {\begin{array}{*{20}{l}}
{0\,\,0}\\
{1\,\,0}\\
{0\,\,1}
\end{array}} \right]
\end{array} \label{eq:133}
\end{align}
}

\begin{figure}
\begin{center}
\includegraphics[height=6 cm]{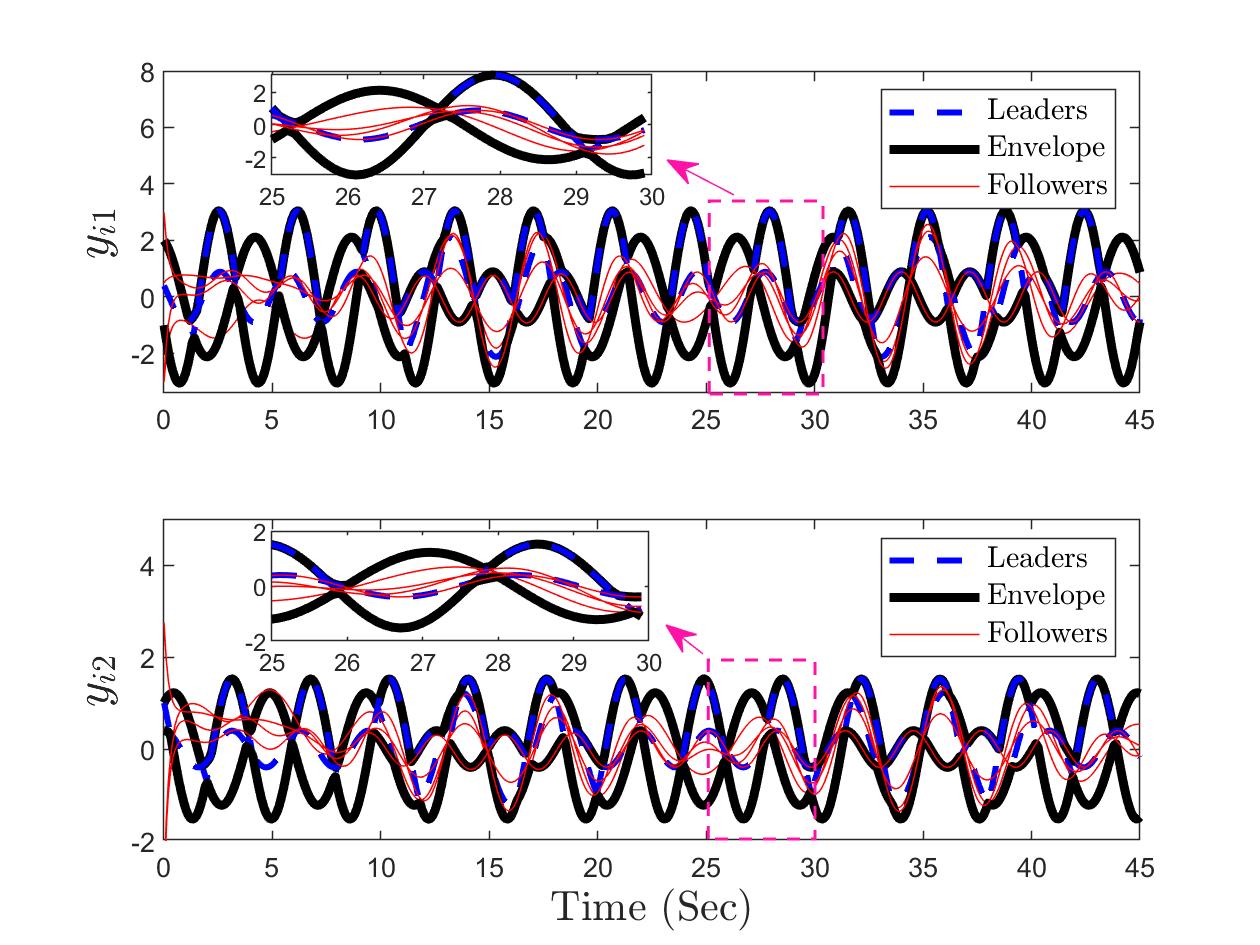}     
\caption{The outputs of all agents.}
\label{fig3}                     
\end{center}                                
\end{figure}

Applying the proposed control law (\ref{eq:118}), Problem~3 with multi-agent systems (\ref{eq:133}) and the given communication graph in Fig.~\ref{fig1} is solved. Fig.~\ref{fig3} shows that the outputs of the followers are converged to the envelopes which formed by the leaders' outputs and stay in them.  These results show that the containment control of MAS is successfully achieved.

 \vspace{-0.1 cm}

\section{Conclusion}
The distributed containment control problem of heterogeneous multi-agent systems with heterogeneous linear dynamics leaders was studied in this paper. We first converted the output containment problem into multiple reference trajectories tracking problem, in which each follower aims to track its virtual exo-system output. To build this virtual exo-system by each follower, a novel distributed algorithm was developed. Novel distributed control protocols were then designed to solve the FHCCP.
The effectiveness of theoretical results was verified by performing numerical simulations. 

\bibliographystyle{IEEEtran}
\bibliography{Main TAC brief V1}

\vfill

\vspace{0.2in}
\end{document}